\documentclass[]{aastex631}

\def\edithere{\textcolor{blue}}

\usepackage{amsmath}
\usepackage{amssymb}
\begin{document}

\title{\uppercase{Pulsed, Polarized X-ray Emission from Neutron Star Surfaces: \\
the Effects of Vacuum Birefringence in the Magnetosphere}}

\author[0000-0001-9268-5577]{Hoa Dinh Thi}
\affiliation{Department of Physics and Astronomy—MS 108, Rice University, 6100 Main St., Houston, TX 77251-1892, USA}


%
%
\font\fiverm=cmr5       \font\fivebf=cmbx5      \font\sevenrm=cmr7     
\font\sevenbf=cmbx7     \font\eightrm=cmr8      \font\eighti=cmmi8     
\font\eightsy=cmsy8     \font\eightbf=cmbx8     \font\eighttt=cmtt8    
\font\eightit=cmti8     \font\eightsl=cmsl8     \font\sixrm=cmr6       
\font\sixi=cmmi6        \font\sixsy=cmsy6       \font\sixbf=cmbx6      
\def\dover#1#2{\hbox{${{\displaystyle#1 \vphantom{(} }\over{
   \displaystyle #2 \vphantom{(} }}$}}

\def\fsc{\alpha_{\hbox{\sevenrm f}}}                                
\def\sigt{\sigma_{\hbox{\fiverm T}}}                                
\def\taut{\tau_{\hbox{\fiverm T}}}                                
\def\lambar{\lambda\llap {--}}
\def\omegaB{\omega_{\hbox{\sixrm B}}}
\def\SigmaB{\Sigma_{\hbox{\sixrm B}}}
\def\DeltaB{\Delta_{\hbox{\sixrm B}}}
\def\wcyc{\omega_{\hbox{\fiverm B}}}
\def\thetaMSC{\theta_{\hbox{\sixrm MSC}}}
\def\thetaB{\theta_{\hbox{\fiverm B}}}
\def\muB{\boldsymbol{\mu}_{\rm B}}
\def\rSch{R_{\hbox{\sixrm S}}}
\def\rns{R_{\hbox{\sixrm NS}}}
\def\mns{M_{\hbox{\sixrm NS}}}
\def\Psis{\Psi_{\hbox{\sixrm S}}}

\def\hruleseparator{\vskip 8pt \textcolor{blue}{\centerline{\vbox{\hrule height 2.0pt width 100pt}}} \vskip 10pt}

\def\teq#1{$\, #1\,$}                         
\def\edithere#1{\textcolor{black}{#1}}  
\def\actionitem#1{\textcolor{blue}{#1}}  
\def\commenthere#1{\textcolor{green}{#1}}  

\author[0000-0003-4433-1365]{Matthew G. Baring}
\affiliation{Department of Physics and Astronomy—MS 108, Rice University, 6100 Main St., Houston, TX 77251-1892, USA}

\author[0000-0002-9705-7948]{Kun Hu}
\affiliation{Physics Department, McDonnell Center for the Space Sciences, and Center for Quantum Leaps, \\
Washington University in St. Louis, St. Louis, MO 63130, USA}

\author[0000-0001-6119-859X]{Alice K. Harding}
\affiliation{Theoretical Division, Los Alamos National Laboratory, Los Alamos, NM 58545, USA}

\author[0000-0002-0254-5915]{Rachael E. Stewart}
\affiliation{Department of Physics, George Washington University 725 21st Street NW, Washington, DC 20052, USA}

\author[0000-0002-7991-028X]{George A. Younes}
\affiliation{Astrophysics Science Division, NASA Goddard Space Flight Center, 8800 Greenbelt Road, Greenbelt, MD 20771, USA}
\affiliation{Center for Space Sciences and Technology, University of Maryland Baltimore County, \\
1000 Hilltop Cir, Baltimore County, MD 21250, USA}

\author[0000-0003-0503-0914]{Joseph A. Barchas}
\affiliation{Department of Astronomy, Geology, and Physics, Southwest Campus, Houston City College, \\
5601 W. Loop S., Houston, TX 77081, USA}



\begin{abstract}
Intense magnetic fields in the atmospheres of neutron stars render non-trivial angular dependence of intensity and polarization of soft X-ray emission originating from their surfaces. By tracking the complex electric field vector for each photon during its atmospheric transport and propagation in general relativistic and birefringent magnetospheres, our Monte Carlo simulation, named {\sl MAGTHOMSCATT}, allows for capturing the complete polarization properties, including the intricate interplay between linearity and circularity.  The new inclusion in {\sl MAGTHOMSCATT} of quantum electrodynamical influences on polarization in the magnetosphere is presented.  We simulate the pulsed and polarized X-ray emission from the outer layers of optically thick, fully ionized atmospheres of neutron stars, with a focus on the radiation emitted from  extended polar caps of magnetars, which are the most highly magnetized neutron stars.   Using the recent intensity pulse profile data for the magnetar 1RXS J1708-4009, we constrain the geometric parameters, namely the angles between the magnetic axis and the observer’s viewing direction relative to the spin axis, as well as the sizes of emission regions. The distributions of these parameters  and the best-fit configuration are provided. In addition, we discuss the important impacts of vacuum birefringence in the magnetosphere on increasing the linear polarization degree. A comparison with the case of a weakly magnetized neutron star,  RX J0822.0-4300, is also discussed.  Our simulation still needs further development, particularly to incorporate the vacuum resonance effect. Nevertheless, the formalism presented here can be employed to constrain geometric parameters for various types of neutron stars.
\end{abstract}



\section{Introduction}
 \label{sec:intro}
Magnetars are the most magnetic members of the isolated neutron star (NS) family, typically possessing extreme fields, \teq{B_p > 10^{14}}Gauss at the surface magnetic pole \citep[e.g.,][]{Kouveliotou-1998-Nature}.  Being bright and hot steady X-ray sources, with luminosities \teq{L_{\rm X,\, 0.5-8~keV}\sim 10^{32}$-$10^{36}} erg/sec usually exceeding their rotational energy losses, they sporadically emit intense bursts in hard X-rays/soft $\gamma$-rays. They typically possess long spin periods, \teq{P\sim 1-12}~s, and large spin-down rates, \teq{\dot{P} \sim 10^{-11}-10^{-13}}~s~s$^{-1}$ \citep{OK-2014-ApJS}, which are ascribed to magnetic dipole torques on the stellar rotation. For vacuum magnetospheres, the \teq{(P, {\dot P})} determinations measure \teq{B_p\sin\alpha}, where \teq{\alpha} is the inclination angle between the magnetic dipole (\teq{\muB}) and rotation (\teq{\boldsymbol{\Omega}}) axes.  The dependence on \teq{\alpha} is much more muted in plasma magnetospheres.  Notwithstanding, magnetar \teq{B_p} estimates are somewhat uncertain, and this imprecision feeds into the quantum and classical physics employed in models that describe their crustal interiors, surfaces and magnetospheres. Reviews of the magnetar landscape can be found in \cite{Turolla-2015-RPPh,KB-2017-ARAA}.

Magnetar persistent X-ray emission at \teq{\lesssim 5}keV is primarily quasi-thermal and is believed to emanate from their hot stellar surfaces, with effective temperatures \teq{kT_{\rm eff}\approx 0.5}~keV \citep{Perna-2001-ApJ,Vigano-2014-MNRAS}. These surface signals are perhaps the most useful tool for probing the equation of state of magnetar interiors. Extending to the full \teq{0.5 - 10}keV window, their soft X-ray spectra are best fit by either a combination of two thermal blackbody components with temperatures around $\sim$0.3~keV (cooler) and $\sim$0.7~keV \citep[hotter, e.g. see][]{Tiengo-2008-ApJ}, or a thermal blackbody component with a temperature around \teq{\sim 0.4}~keV, plus a steep non-thermal power-law component \teq{E^{-\Gamma}} with index \teq{\Gamma \gtrsim 2} \citep[e.g.,][]{Zane-2023-ApJ}.  The pulse profile of \teq{E < 5}keV emission is usually broad and sometimes single-peaked as it is for 1RXS J1708-4009 \citep{Zane-2023-ApJ} and 1E~1547-5408 \citep{Bernardini-2011-AandA}. Other magnetars such as 4U~0142+0162 exhibit a double-peaked light curve \citep{Rea-2007-MNRAS} with a small pulse fraction \teq{\sim 10-20}\% below 10 keV, a possible indicator that a large portion of the surface is emitting.  A pathological example is the time-evolving, multi-peaked profile \citep{Younes-2022-ApJ} of a recently discovered magnetar SGR 1830-0645.  There is also the appearance of additional peaks at energies \teq{E \gtrsim 5}keV, such as for 1RXS J1708-4009 \citep{denHartog-2008-AandA,Zane-2023-ApJ}, which may constitute a coronal emission component coming from just above the surface.

Soft X-ray pulsation is the telltale signature of non-uniform emission across a NS surface.  This feature has been exploited for magnetars by several groups to probe the size of the hot emission region, the value of the star's magnetic inclination angle \teq{\alpha} in a dipolar field morphology, and the observer's viewing direction angle \teq{\zeta} relative to rotation axis \teq{\boldsymbol{\Omega}}.  For instance, \cite{Bernardini-2011-MNRAS,Albano-2010-ApJ} fit the pulse profiles of XTE~J1810-197 during quiescence while \citep{Albano-2010-ApJ} did the same for CXOU~J164710.2-455216 (finding \teq{\alpha\sim 80^{\circ}} and \teq{\zeta\sim 25^{\circ}}), employing blackbody emission from the whole surface with a prescribed colatitudinal temperature distribution. Both works included gravitational light bending.  With a similar construct, \cite{Guillot-2015-MNRAS} found that a very small polar region with a temperature a factor of 6 larger than for an equatorial zone is required to fit the single-peaked soft X-ray profile for the so-called low-field magnetar SGR~0418+572.  More recently, \cite{Younes-2020-ApJ} fit the double-peaked pulse profile of a 2.2 hour flare detected from 1RXS J1708-4009 with \teq{\alpha\sim 60^{\circ}} and \teq{\zeta\sim 60^{\circ}}, concluding that antipodal spots are required to match the flare light curve: this implies that the flare heating must be connected to the global dipolar fields of the star. Omitted from these and other similar studies are detailed treatments of local radiation anisotropy, which varies in atmospheres as the direction and magnitude of $\boldsymbol{B}$ changes across the stellar surface.

The uncertainties in inferred geometrical stellar parameters \teq{\alpha} and \teq{\zeta} are considerable when only modeling intensity pulse profiles. This situation can potentially be ameliorated if pulsed polarimetric data is also available; this is now possible for magnetars.  In the last three years, the Imaging X-ray Polarimetry Explorer ({\sl IXPE}), sensitive in the \teq{2-8}~keV band, has delivered key polarization results so far for 6 magnetars, with a notable diversity of polarization signatures. For 1RXS J1708-4009, the {\sl IXPE} pulse profile is essentially single-peaked at \teq{2-4}keV, and the polarization degree (PD) is anti-correlated with the intensity \citep[see][]{Zane-2023-ApJ}.  In addition, the polarization angle (PA) on the sky associated with the Stokes parameters was the same at all energies. This ``fixed'' PA strongly contrasts the wide-ranging {\sl IXPE}~PA values seen in 4U 0142+61 between \teq{<4}keV (where the PD is 14\%) and \teq{>5.5}keV (the PD is 42\%) energies \citep{Taverna-2022-Sci}.   For this magnetar, the observed \teq{90^{\circ}} PA dichotomy perhaps can be interpreted as the \teq{>5.5} keV signal emanating from a coronal region where resonant cyclotron scattering of surface photons occurs  \citep{Taverna-2022-Sci}. 

High fidelity, phase-resolved {\sl IXPE} polarimetry for 1E 1841-045 revealed a limited variation of both the PD (at around \teq{\sim 20-25}\%) and PA, and a prominent power-law component above around 5 keV \citep{Stewart-2025-ApJL,Rigoselli-2025-ApJ} that almost certainly originates in the inner magnetosphere. SGR 1806-20 provided a further puzzle, with significant PD levels of 40\% only in the $4-5$ keV window, and less than 5\% below 4 keV, leading to the suggestion \citep{Turolla-2023-ApJ} that the thermal emission may originate in a condensed part of the star's surface.  PD values below 25\% were obtained for 1E 2259+586 \citep{Heyl-2024-MNRAS}, again being attributed to a condensed surface, a state that might be expected \citep{Medin-2007-MNRAS} for some of a magnetar's parameter space. An alternative explanation for the low PDs and energy-dependent PAs \citep{Lai-2023-PNAS} could be signatures of the physics of the vacuum resonance, a density/frequency feature that results from anomalous dispersion when plasma and vacuum birefringence are in strong competition with each other deeper in a magnetar atmosphere.   Partial switching between the two linear polarization modes naturally occurs \citep{Lai-Ho-2002-ApJ}, leading to a depolarization of the signal emergent from magnetar surfaces.

To help resolve the debate on whether magnetars possess atmospheres or condensed surfaces, or both, it is necessary to have a versatile outer atmosphere model at our disposal, one that can address a wide variety of locales on the stellar surface, and not just at the magnetic pole. Towards this goal, and to accommodate a similar objective for pulsars of lower magnetizations, our group has developed a Monte Carlo simulation, {\sl MAGTHOMSCATT}, which tracks polarized photon propagation and scattering in the magnetic Thomson domain, in atmospheric slabs threaded by $\boldsymbol{B}$ fields in arbitrary directions.  The developmental technical details are described in \cite{Barchas-2021-MNRAS,Hu-2022-ApJ}, and also in \cite{Dinh-2025-ApJ}, in which algorithmic upgrades that lead to improved efficiency in the magnetar domain are described. Sample results that include the influence of general relativity on propagation of light to infinity can be found in \cite{Hu-2022-ApJ,Dinh-2025-ApJ}.  A brief summary of its construction is provided in Section~\ref{sec:radiative-transfer}.  {\sl MAGTHOMSCATT} fully addresses light bending and parallel transport of photon electric field vectors in the Schwarzschild metric, the non-rotating general relativistic geometry that is appropriate for magnetars.  Yet it has not directly addressed the modifications to polarized photon transport imbued by the birefringent vacuum in the magnetosphere.  This is the development we put forward in this paper, with a detailed exposition commensurate with the prior works of \citep{Heyl-2000-MNRAS, Adelsberg-2006-MNRAS} provided in Section~\ref{sec:propagation}.  A focal study of the pulsed intensity and polarization characteristics of the magnetar 1RXS J1708-4009 is offered in Section~\ref{sec:result-and-discussion}, wherein a large increase in PD is imposed by birefringence of the magnetosphere.  The impacts of birefringence in a NS of lower magnetization, specifically RX~J0822.0-4300, are shown to be significant, albeit not as pronounced.  {\sl MAGTHOMSCATT} continues to be upgraded, and our program will address the vacuum resonance physics inside the atmosphere and its potential to depolarize the surface signals in a future work.

\section{Monte Carlo Simulations of  Radiative Transfer in Neutron Star Atmospheres}
\label{sec:radiative-transfer}
In this work, we model X-ray signals from NS surfaces using {\sl MAGTHOMSCATT}, a Monte Carlo simulation that models \edithere{transport and diffusion in ionized outer atmospheres and tracks photon complex electric field vectors $\boldsymbol{\mathcal{E}}$, whose 3D distribution is modified by Thomson scatterings in strong magnetic fields; see \cite{Barchas-2021-MNRAS} for details of the differential cross section, first derived by \cite{Canuto-1971-PRD}.}  The original design of the simulation is detailed in \citet{Barchas-2021-MNRAS}, while recent developments can be found in \citet{Hu-2022-ApJ} and \citet{Dinh-2025-ApJ}.  In the following, we summarize the key ingredients of {\sl MAGTHOMSCATT} in Section~\ref{subsec:magthomscatt} and the main elements  pertinent to magnetar simulations  \citep{Dinh-2025-ApJ} in Section~\ref{subsec:magnetar}.

\subsection{MAGTHOMSCATT: Summary of Key Elements}  
 \label{subsec:magthomscatt}
The position of an emission locale on a NS surface can be defined by its  magnetic colatitude $\theta$ and longitude $\phi$.  
For  X-ray emission from a spherical rectangular patch on the surface, the emission locales within this emitting region are uniformly sampled according to:
\begin{equation}
	\cos \theta  \; = \; \cos \theta_l + (\cos \theta_u - \cos \theta_l) \, \Xi_\theta \quad , \quad \phi \; = \; \phi_l + (\phi_u - \phi_l) \, \Xi_{\phi} \quad , \label{eq:emission_locale}
\end{equation}
where $\left[\theta_l, \theta_u \right]$ and $\left[\phi_l, \phi_u\right]$ are respectively the intervals of magnetic colatitudes and longitudes of the patch.
The two random variables, $\Xi_{\theta}$ and $\Xi_{\phi}$, are generated between between 0 and 1. 
Throughout this work, we limit our analyses to emission from azimuthally symmetric polar caps. Therefore, $\phi \in \left[0^{\circ}, 360^{\circ}\right]$, and $\theta \in \left[0^{\circ}, \theta_{\rm cap}\right]$ ($\theta \in \left[180^{\circ} - \theta_{\rm cap}, 180^{\circ}\right]$) if the cap is located at the ``north'' (``south'') pole, with $\theta_{\rm cap}$ being the angular size of the cap.

Each surface locale can be represented by a slab geometry; see Figure~1 in \citet{Barchas-2021-MNRAS} for an illustration.  The transport properties of  photons within the slab are dependent on the frequency ratio $\omega /\wcyc$ -- where $\omega$ and  $\wcyc = eB / (m_ec)$ are respectively the photon and electron cyclotron frequencies, with $B = |\boldsymbol{B}|$ being the magnetic field strength and $e$, $m_e$, $c$ respectively denoting the electron charge, electron mass,  and the speed of light -- and the orientation of the  local magnetic field vector $\boldsymbol{B}$ with respect to the slab normal vector $\boldsymbol{\hat{n}}$. The latter can be characterized by the angle $\theta_{\rm B} = \arccos\left(\boldsymbol{\hat{n}} \cdot  \hat{\boldsymbol{B}} \right)$, where $\hat{\boldsymbol{B}} = \boldsymbol{B} /B$ is the unit magnetic field vector. For any symmetric magnetic field configuration, 
  $\theta_{\rm B} = 0^{\circ}$ ($\theta_{\rm B} = 90^{\circ}$)  at the 
pole (equator).  In the simulation, photons are injected at the bottom the slab and  then undergo magnetic Thomson scatterings. Photons that end up coming out from the base cannot be recorded by observers, and thus, only those that emerge from the upper boundary of the surface layer are of interest.  Once they successfully escape the NS surface, photons propagate through the magnetosphere, ultimately reaching observers at infinity; see Figure 1 in \citet{Hu-2022-ApJ} for an illustration of the global geometry. We will present details of magnetospheric propagation, including general relativistic (GR)  and quantum electrodynamical (QED) effects, in Section~\ref{sec:propagation}. In this Section, we will solely focus on describing simulation elements within the atmospheres.

As previously mentioned, our simulation adopts an electric field vector formalism, where the complex electric field vector $\boldsymbol{\mathcal{E}}$ of each photon is tracked during its scattering in the NS atmosphere and propagation in the magnetosphere.  This approach offers the flexibility of projecting the electric field vector onto a coordinate system defined by the photon propagation vector and any reference direction. As a result, the Stokes parameters, $I, Q, U, V$, in any coordinates can easily be calculated. Here, $I$ is the radiation intensity, $Q, U$ are related to linear polarization information,
and $V$ corresponds to circular polarization. This versatile feature of {\sl MAGTHOMSCATT}  sets it apart from ealier approaches that relied on tracking Stokes parameter information \citep[see, e.g.,][]{Whitney1991} or solving the radiative transfer equation for two normal modes \citep[see, e.g.,][]{Ho2001,Ozel2001}. 
 {\sl MAGTHOMSCATT} was thoroughly validated by various direct comparisons with  results in the literature. In particular, \citet{Barchas-2021-MNRAS} demonstrated that the intensity and polarization distributions obtained with {\sl MAGTHOMSCATT} at different sub-cyclotronic and super-cyclotronic frequency ratios, $\omega/\wcyc = 0.25, 0.5, 2, 10$, are consistent with those  obtained by \cite{Whitney1991}. In addition, \citet{Dinh-2025-ApJ} recently performed a simulation at $\omega / \wcyc = 1000$ to approximate the non-magnetic domain and showed that numerical solutions to radiative transfer integro-differential equation by \cite{Chandrasekhar-1960-book} and \cite{Sunyaev-1985-AandA} are reproduced with {\sl MAGTHOMSCATT}.

Regarding the intensity and polarization properties of photons at the point of injection, we employ the so-called anisotropic and polarized (AP) protocol proposed in \citet{Barchas-2021-MNRAS}, and honed in \citet{Hu-2022-ApJ} and \citet{Dinh-2025-ApJ}. The AP injection approach was shown to be more efficient than the isotropic and  unpolarized (IU) counterpart, particularly so for the magnetar domain; see Table~3 in \citet{Baring-ApJ-2025} for the run time ratios between AP and IU at different frequency ratios and $\theta_{\rm B}$ values. The anisotropy of photons at injection is generated  with respect to the magnetic field direction using the accept-reject method  according to the following distribution:
\begin{equation}
	I(\mu_0) \; =\; \frac{A_{\omega}(\mu_0)}{\Lambda_{\omega} \sigma(\omega, \mu_0)}
	\quad ,\quad
	\Lambda_{\omega} \; =\; 
	\int_{-1}^{1}\frac{A_{\omega}(\mu_0)}{\sigma(\omega, \mu_0)} d\mu_0 \quad , \quad
	A_{\omega}(\mu_0)  \; = \; \frac{3}{2} \frac{1+ \mathcal{A}(\omega) \mu_0^2}{3+ \mathcal{A}(\omega)}  \quad ,
	\label{eq:I_mui}
\end{equation}
where $\mu_0 = \hat{\boldsymbol{k}}_0 \cdot  \hat{\boldsymbol{B}}$ is the cosine of the angle between the injected propagation direction $\hat{\boldsymbol{k}}_0$ and the magnetic field direction, and $\sigma(\omega, \mu_0)$ is the total magnetic Thomson scattering cross section; see Equation~(13) in \citet{Barchas-2021-MNRAS}. The photon electric field vector  at injection, $\boldsymbol{\mathcal{E}}_0$, can be written in terms of two components  $\mathcal{E}_{\theta}$ and $\mathcal{E}_{\phi}$ as:
\begin{equation}
	\boldsymbol{\mathcal{E}}_0  \; =\; \mathcal{E}_{\theta} \hat{\theta} + \mathcal{E}_{\phi} \hat{\phi}
	\quad  
	\hbox{with}\quad
	\hat{\phi} \; =\; \frac{ \hat{\boldsymbol{B}} \times \hat{\boldsymbol{k}}_0}{|\hat{\boldsymbol{B}} \times \hat{\boldsymbol{k}}_0|}  \quad , \quad  \hat{\theta}  = \hat{\phi} \times \hat{\boldsymbol{k}}_0 \ , \nonumber
	\label{eq:E_sph-system}
\end{equation}
and $\mathcal{E}_{\theta}$  and $\mathcal{E}_{\phi}$  can be written  in terms of the two Stokes parameters $\hat{Q}_{\omega}$ and $\hat{V}_{\omega}$ \citep{Barchas-2021-MNRAS}:
\begin{equation}
	\mathcal{E}_{\theta} \; =\; \sqrt{\frac{\Pi_{\omega} + \hat{Q}_{\omega}}{2\Pi_{\omega} }}
	\quad , \quad \mathcal{E}_{\phi} \; =\; \frac{ i \hat{V}_{\omega}}{2\Pi_{\omega}	\mathcal{E}_{\theta}} \quad , \quad \Pi_{\omega}  = \sqrt{\hat{Q}^2_{\omega} + \hat{V}^2_{\omega}} \quad ,
	\label{eq:E_theta_phi}
\end{equation}
\begin{equation}
	\hat{Q}_{\omega}(\mu_0) \; =\; \frac{\mathcal{A}(\omega)[\mu_0^2 - 1]}{1+ \mathcal{A}(\omega)\mu_0^2} 
	\quad , \quad 
	\hat{V}_{\omega}(\mu_0) \; =\; \frac{2\mathcal{C}(\omega)\mu_0}{1+ \mathcal{A}(\omega)\mu_0^2} 
	\quad .
	\label{eq:Stokes-inj}
\end{equation}
The coefficients $\mathcal{A}(\omega)$ and $\mathcal{C} (\omega)$ in the expressions of $I(\mu_0)$, $\hat{Q}_{\omega}(\mu_0) $, and $\hat{V}_{\omega}(\mu_0)$ pertain to the anisotropy and circular polarization and are given by Equations~(35) and (36) in \citet{Baring-ApJ-2025}.

The thickness of an atmospheric slab is defined via 
an effective optical depth parameter $\tau_{\rm eff}$, which depends on the magnetic field direction and relates to the Thomson optical depth \teq{\taut}; see Equation~(5) in \citet{Hu-2022-ApJ}. The optimal value of $\tau_{\rm eff}$ corresponds to the minimum value of $\tau_{\rm eff}$ at which convergence is reached. In general, the optimal $\tau_{\rm eff}$ value varies with $\theta_{\rm B}$ and $\omega / \wcyc$.  As shown in Table~3 in \cite{Baring-ApJ-2025}, for $\omega / \wcyc \sim 1$ and $\omega / \wcyc > 1$, frequencies that are relevant to X-ray emission from NSs of low magnetizations, such as central compact objects (CCOs),
an effective optical depth of $\tau_{\rm eff} = 6$ is sufficient across all magnetic colatitudes.
In the magnetar domain, where $\omega / \wcyc \ll 1$, \edithere{such a single $\tau_{\rm eff}$ choice is not appropriate,} and so we incorporate on top of the AP injection protocol a prescription for the minimum number of scatterings. This new approach, named AP*, was recently introduced in \citet{Dinh-2025-ApJ}, and is  described below in the next Section.  Note that our simulation applies to the outer layers of an atmosphere; contributions from free-free opacity that arise deeper in atmospheres where the densities are higher will be introduced in a future upgrade of the simulation.

\subsection{Simulation Approach in the Magnetar Domain}
 \label{subsec:magnetar}
In the magnetar regime, where $\omega / \wcyc \ll 1$, the AP injection significantly improves the simulation efficiency with respect to the IU counterpart. In particular, for emission locales near the equator, the simulation run time is reduced by three orders of magnitude, as shown in \cite{Baring-ApJ-2025}. However, due to the pronounced disparity in the scattering cross sections between different polarization modes and propagation directions in this domain, the simulation requires longer run time to achieve convergence than other frequency domains, for the same statistics. Indeed, as discussed in \citet{Barchas-2021-MNRAS}, at substantially 
sub-cyclotronic frequencies, the differential cross section of photons in the perpendicular ($\perp$) mode scales as $(\omega/\wcyc)^2$. On the contrary,  the cross section for the parallel ($\parallel$) polarization mode retains a frequency-independent component that becomes dominant at low $\omega / \wcyc$; see Equation~(B9) in \citet{Barchas-2021-MNRAS}. Consequently, except when the photon incident angle to the magnetic field direction drops to below approximately $\theta_{\rm MSC} = \omega/\wcyc$, where the cross sections for both polarizations become comparable, the characteristics of the polarized magnetic Thomson transport in the atmosphere remain independent of the frequency ratio. Here, $\theta_{\rm MSC}$ is the opening angle of the so-called magnetic scattering cone (MSC) \citep{Baring-ApJ-2025}.
As a result, within the MSC, the radiation is strongly beamed along the magnetic field and becomes depolarized, whereas outside this cone, it is highly polarized in the $\parallel$ mode.

To achieve these characteristics of radiation more efficiently, in \cite{Dinh-2025-ApJ} we combined the AP injection protocol with the following prescription for the minimum number of scatterings for each photon:
 \begin{equation}
 	n_{\rm scat, min } \quad = \quad
 	\begin{cases} 
 		5H(\theta_i -2\thetaMSC) & \text{if $\theta_{\rm B} \leq 2^{\circ}$}  \\
 		5H(\theta_i -0.1) & \text{if $\theta_{\rm B} > 2^{\circ}$}
 	\end{cases} \quad ,
 	\label{eq:nscat_min}
 \end{equation}
where $\theta_i$ is the photon incident angle (in radians) with respect to the magnetic field direction, and $H(x)$ is the Heaviside step function, with $H(x) = 1$ if $x\geq 0$, and zero otherwise. Thus, the minimum number of scatterings of each photon depends on its propagation direction with respect to the magnetic field direction as well as its frequency relative to the electron cyclotron frequency. This new protocol, AP*, was shown to reduce the run time by a factor of $\sim 2 -8$ compared to the standard AP, where minimum scattering number is fixed at $n_{\rm scat, min} = 1$; see Table~1 in \citet{Dinh-2025-ApJ}.

The properties of the cross sections discussed above indicates that the behavior of the atmospheric radiative transfer is only influenced by the frequency ratio within a narrow  cone around the field direction. This cone becomes negligibly small for $\omega / \wcyc \lesssim 0.01$, where $\theta_{\rm MSC} < 0.57^{\circ}$. For emission from extended polar caps, the regions around the poles where the magnetic cone has a significant impact constitute a small fraction of the cap surface, hence contributing only a minority of the overall signal. Accordingly, once $\omega / \wcyc$ becomes sufficiently small, the emission signatures of anisotropy and polarization  remain the same as $\omega / \wcyc$ further decreases. Indeed, in our recent paper \cite{Dinh-2025-ApJ}, we showed that the intensity and polarization pulse profiles become essentially invariant with \teq{\omega / \wcyc} when this ratio drops below 0.01. For this reason,  to model X-ray emission from extended surface regions of magnetars, we performed the simulation at   $\omega/\wcyc  = 0.01$ in the local inertial frame (LIF) at the magnetic poles. This approach significantly reduces the run time while still maintaining the robustness of the simulation.

\section{Propagating Neutron Star Surface Emission to Infinity}
 \label{sec:propagation}
The anisotropy and polarization characteristics of soft X-ray emission from NS surfaces observed at infinity are determined not only by the scattering process in the atmosphere but also by the propagation of photons through the magnetosphere.  
The Stokes parameters at any altitude are calculated using Cartesian coordinates with basis vectors defined by the unit propagation wavevector $\hat{\boldsymbol{k}}_{\rm GR}$ and unit neutron-star rotation vector $\hat{\boldsymbol{\Omega}}$ as follows
\begin{equation}
	\hat{z} \; = \;  \hat{\boldsymbol{k}}_{\rm GR} \quad , \quad   \hat{y} \; = \; \frac{\hat{\boldsymbol{\Omega}} \times \hat{\boldsymbol{k}}_{\rm GR}}{|\hat{\boldsymbol{\Omega}}  \times \hat{\boldsymbol{k}}_{\rm GR}|} \quad , \quad \hat{x} \; = \;  \hat{y} \times \hat{\boldsymbol{k}}_{\rm GR} \quad . \label{eq:coordinate}
\end{equation}
The photon electric field vector measured at altitude \teq{r} can be decomposed into two components:
\begin{equation}
	\boldsymbol{\mathcal{E}}_{\rm GR} \; = \; \mathcal{E}_{x} \hat{x} +  \mathcal{E}_{y} \hat{y} \quad ,
 \label{eq:E_infinity_def}
\end{equation}
and Stokes parameters can be obtained with $ \mathcal{E}_{x}$ and $ \mathcal{E}_{y}$ using Equation~(A4) in \citet{Hu-2022-ApJ}. At large altitudes, the photon propagation is radial and one sets \teq{\hat{\boldsymbol{k}}_{\rm GR} \to \hat{\boldsymbol{k}}_{\infty} = \hat{r}} in Equation~(\ref{eq:coordinate}) to establish the coordinate description at infinity.  The corresponding photon electric field vector pertinent to a distant observer is \teq{\boldsymbol{\mathcal{E}}_{\rm GR} \to \boldsymbol{\mathcal{E}}_{\infty}}.  This Section will focus on describing the modeling of magnetospheric propagation of photons in our {\sl MAGTHOMSCATT} simulation.

As photons leave the NS surface and travel toward a distant observer, their trajectories are curved due to light bending in the  surrounding strong gravitational field. This allows more visibility of the hot spots, hence reducing intensity variation. The impact of general relativity on a photon's four momentum can be characterized by the parameter $\Psi = \rSch/r$, where $r$ is the distance from the photon to the NS center, and $\rSch = 2G\mns /c^2$ is the Schwarzschild radius, with $G$ being the gravitational constant and $M_{\rm NS}$ being the NS mass.  At the surface, $\Psi \to \Psis = 2G\mns /(c^2\rns )$ represents the compactness of the star, with $R_{\rm NS}$ being the NS radius. Throughout this paper, we set $\mns = 1.44 M_{\odot}$ and $\rns = 10$ km, corresponding to a compactness of $\Psis \approx 0.425$.  As the implementation of light bending and parallel transport in {\sl MAGTHOMSCATT} has been detailed in the prior papers of \cite{Barchas-2021-MNRAS} and \cite{Hu-2022-ApJ}, we summarize the treatment of GR effects in Appendix~\ref{sec:GRprop}.

\subsection{Vacuum birefringence in the magnetosphere}
 \label{sec:VB_basics}
In the presence of the external magnetic field, the vacuum becomes birefringent; that is, the vacuum acquires two different refractive indices for the two photon propagation eigenmodes, except for the propagation along the field direction \citep{Adler1970, Adler1971,Tsai-1975-PhRvD}. These indices of refraction can be determined from the Heisenberg-Euler effective Lagrangian of the electromagnetic field in the limit of low frequencies \citep{Adler1971, Heyla1997, Heylb1997}, or from the Kramers-Kronig relations, leveraging the absorptive (imaginary) contribution from single photon pair production \citep{Adler1970, Adler1971}.  Despite QED being one of the most rigorously tested theories in physics, direct measurements  of vacuum birefringence (VB) in laboratories remain an experimental challenge; see, e.g., \cite{Ejlli2020, Zavattini2022}, and references therein. As the vacuum surrounding NSs is highly magnetized, it is expected to create ideal conditions for VB to have significant impacts, particularly altering the polarization properties of photons propagating through the magnetosphere. Therefore, polarization measurements of NS surface emission could serve as an alternative approach to probing such a prediction of QED.

In the modest/weak field domain, the difference in the refractive indices corresponding to the two polarization modes is proportional to the square of the magnetic field strength, as shown in Equation~(6) in \cite{Heyl-2000-MNRAS}:
\begin{equation}
	\Delta n \; = \; \dover{\fsc}{30\pi} \left(\frac{B}{B_{\rm cr}}\right)^2\sin^2 \theta_{\rm kB} 
    \quad , \quad
    B \;\lesssim\; B_{\rm cr}
 \label{eq:deltan}
\end{equation}
with \teq{\fsc} being the fine-structure constant, $B_{\rm cr} \approx 4.4 \times 10^{13}$ Gauss being the quantum critical field strength, and $\theta_{\rm kB}$ the angle between the direction of propagation (${\bf{k}_{\rm GR}}$) and the external field $\boldsymbol{B}$.  The dependence of \teq{\Delta n} on \teq{\vert \boldsymbol{B}\vert} weakens \citep{Tsai-1975-PhRvD,Heyl-2000-MNRAS} in the supercritical field domain, \teq{B> B_{\rm cr}}, which is realized near magnetar surfaces.   Yet, even so, close to the stellar surface where the magnetic field is strong, \teq{\Delta n} is not very small and the two polarization modes are decoupled and evolve independently. As the photon propagates away from the surface, the magnetic field strength quickly drops as $|\boldsymbol{B}| \propto r^{-3}$; see Equation~(\ref{eq:BGR}).  Accordingly, the impacts of VB decline and eventually vanish, and the two polarization modes then recouple at moderate or high altitudes (yet still well within the light cylinder radius), depending on the strength of the surface polar field.  In the following, we present in detail the polarization evolution taking into account the influence of VB in the magnetosphere (Section~\ref{sec:evolution}) and the calculation of the recoupling radius $r_{\rm rec}$ (Section~\ref{sec:recoupling}).

\subsection{Polarization evolution equations}
\label{sec:evolution}

In flat spacetime, there are two main approaches to describing the evolution of polarization in magnetar magnetospheres due to the influences of birefringence of the magnetized QED vacuum, and these will be considered sequentially. As {\sl MAGTHOMSCATT} delivers information on the electric field vector from photons as its output, we both emphasize and focus first on the approach that tracks the evolution of $\boldsymbol{\mathcal{E}}_{\rm GR}$ during propagation.  The electric field vector at any point on the trajectory of a photon can be written in terms of the two polarization modes:
\begin{equation}
	\boldsymbol{\mathcal{E}}_{\rm GR} \; = \; \mathcal{E}_{\rm O} \hat{e}_{\rm O} + \mathcal{E}_{\rm X} \hat{e}_{\rm X} \quad  , \quad 
	\hat{e}_{\rm X} \; = \; \frac{ \boldsymbol{B} \times \hat{\boldsymbol{k}}_{\rm GR}  }{|\boldsymbol{B} \times \hat{\boldsymbol{k}}_{\rm GR}|} \quad , \quad 
	\hat{e}_{\rm O}  \; = \;  \hat{e}_{\rm X} \times \hat{\boldsymbol{k}}_{\rm GR} \quad .
	\label{eq:Evec}
\end{equation}
with $\hat{\boldsymbol{k}}_{\rm GR}$ and $\boldsymbol{B}$ being given in Equations~(\ref{eq:ray-tracing}) and (\ref{eq:BGR}), respectively. In a strongly-magnetized vacuum with field strengths \teq{B\gtrsim B_{\rm cr}}, at X-ray energies the O (X) mode corresponds approximately to the linear polarization state where $\boldsymbol{\mathcal{E}}_{\rm GR}$ is in (perpendicular to) the plane of $\hat{\boldsymbol{k}}_{\rm GR}$ and $\boldsymbol{B}$.   As $\hat{e}_{\rm O}$ and  $\hat{e}_{\rm X}$ rotate with variations in $\boldsymbol{B}$ and $\hat{\boldsymbol{k}}_{\rm GR}$ as the photon propagates through the magnetosphere, $\mathcal{E}_{\rm O}$ and  $\mathcal{E}_{\rm X}$ evolve as \citep{Adelsberg-2006-MNRAS}:
\begin{equation}
	i\begin{pmatrix}
		\dover{d\mathcal{E}_{\rm O}}{ds} \\[4pt]
		\dover{d\mathcal{E}_{\rm X}}{ds}
	\end{pmatrix}
\; \approx \;
 \begin{pmatrix}
 	\dover{-\omega \Delta n}{2c} & i\dover{d\Phi_{\rm B}}{ds} \\[6pt]
 	-i\dover{d\Phi_{\rm B}}{ds}  & \dover{\omega \Delta n}{2c}
 \end{pmatrix}
\begin{pmatrix}
	\mathcal{E}_{\rm O} \\[4pt]
	\mathcal{E}_{\rm X} \\
\end{pmatrix}
\quad ,  
 \label{eq:dE_ds}
\end{equation}
where $s$ is the distance along the direction of propagation.  Herein $\Delta n = n_{\rm O} - n_{\rm X}$ is the difference in the refractive indices between the two modes, given  by Equation~(\ref{eq:deltan}). Also, $\Phi_{\rm B}$ denotes the angle between the component of the external field perpendicular to the propagation direction $\hat{\boldsymbol{k}}_{\rm GR} $ (or, equivalently, $\hat{e}_{\rm O}$ as defined in Equation~(\ref{eq:Evec}) above) and the local x-axis defined by $\hat{\boldsymbol{k}}_{\rm GR}$ and the spin axis $\hat{\boldsymbol{\Omega}}$ in Equation~(\ref{eq:coordinate}).  Thus,
\begin{equation}
	\cos \Phi_{\rm B} \; = \; \hat{e}_{\rm O} \cdot \hat{x} 
	\; =\; \frac{ \boldsymbol{B} \times \hat{\boldsymbol{k}}_{\rm GR}  }{|\boldsymbol{B} \times \hat{\boldsymbol{k}}_{\rm GR}|}
	\cdot \frac{\hat{\boldsymbol{\Omega}} \times \hat{\boldsymbol{k}}_{\rm GR}}{|\hat{\boldsymbol{\Omega}}  \times \hat{\boldsymbol{k}}_{\rm GR}|}
	\;\equiv\; \hat{e}_{\rm X} \cdot \hat{y} 
	\label{eq:PhiB_def}
\end{equation}
defines the ``tilt angle'' $\Phi_{\rm B}$ for the polarization propagation geometry through the magnetosphere.  Observe that a standard vector identity for quadruple cross products has been employed to simplify the vector algebra and thereby identify an algebraically simpler and alternative definition of this angle.

The inherent mixing of modes implied by Equation~(\ref{eq:dE_ds}) indicates that the electric field vectors rotate about their wavevectors as they propagate. The mathematical structure of Equation~(\ref{eq:dE_ds}) combined with the rapid decline of the field strength with altitude (i.e., radius \teq{r}) leads naturally to a bifurcation of the propagation character, separated by a recoupling radius \teq{r_{\rm rec}} (also known as the polarization-limiting radius). Inside this recoupling radius when $r < r_{\rm rec}$, as $\omega \Delta n/(2c) \gg |d\Phi_{\rm B} / ds |$, the off-diagonal terms in Equation~(\ref{eq:dE_ds}) are then negligible.  As a result, the two polarization modes evolve independently and their phases change at a rate of the order of \teq{\omega \Delta n/2} and in an opposite sense. Outside the recoupling radius where $\omega \Delta n/(2c) \ll | d\Phi_{\rm B} / ds |$, the modes are tightly coupled so that their evolution is described by the variation of the \teq{\hat{\boldsymbol{k}}_{\rm GR}-\boldsymbol{B}} geometry along the path. By integrating Equation~(\ref{eq:dE_ds}), at the recoupling radius, the complex amplitudes of the two polarization modes can be expressed as:
%
\begin{equation}
	\mathcal{E}_{\rm O, rec} \; = \;  \mathcal{E}_{\rm O, S}\exp{\left(\frac{i}{2} \Delta \phi \right)} 
	\quad ,\quad  
	\mathcal{E}_{\rm X, rec} \; = \; \mathcal{E}_{\rm X, S}\exp{\left(-\frac{i}{2} \Delta \phi\right)} 
	\qquad \hbox{with}\qquad
	\Delta\phi \; =\; \dover{\omega}{c}\int_0^{s_{\rm rec}} \Delta n \, ds \;\; .
 \label{eq:EX_rc}
\end{equation}
Here, \teq{\mathcal{E}_{\rm O, S}} and \teq{\mathcal{E}_{\rm X, S}} are the complex mode amplitudes at the surface, \teq{\Delta \phi} is the phase shift between the two modes, and $s_{\rm rec}$ is the distance traveled by the photon from the surface to the point of recoupling, and is generally of the order of $r_{\rm rec}$. After the two modes recouple, the electric field vector is parallel-transported to infinity, as described in Section~4 in \citet{Hu-2022-ApJ}. Representative values of the recoupling radius will be outlined in Section~\ref{sec:recoupling}.  We note that for the purposes of the results presented here, up to the recoupling radius, the vacuum birefringence changes to the photon electric field vectors dominate those incurred by parallel transport in the Schwarzschild metric.  Accordingly, we implement a flat spacetime polarization evolution when \teq{\rns < r < r_{\rm rec}}, and only implement parallel transport at \teq{r \geq r_{\rm rec}}.  Yet, \underline{at all radii}, the general relativistic ray tracing outlined in Appendix~\ref{sec:GRprop} is performed.

At the recoupling radius (which lies on a recoupling surface), the Cartesian coordinates defined by the photon propagation direction \teq{\hat{\boldsymbol{k}}_{\rm GR}} and the spin axis \teq{\hat{\boldsymbol{\Omega}}} in Equation~(\ref{eq:coordinate}) lead to the following expressions of the Stokes parameters ($I,Q,U,V$):
%
\begin{eqnarray}
	I & = &  |\mathcal{E}_{\rm O, S}|^2 + |\mathcal{E}_{\rm X, S}|^2  \quad ,    \nonumber  \\[3pt] 
    Q & = & (|\mathcal{E}_{\rm O, S}|^2 - |\mathcal{E}_{\rm X, S}|^2) \cos2\Phi_{\rm B} - 2\left[\Re(\mathcal{E}_{\rm O, S}\mathcal{E}^{*}_{\rm X, S})\cos\Delta\phi -\Im(\mathcal{E}_{\rm O, S}\mathcal{E}^{*}_{\rm X, S})\sin\Delta \phi\right]\sin 2\Phi_{\rm B}  \quad , \nonumber  \\[-5.5pt]
 \label{eq:IQUV_rec} \\[-5.5pt]
	U & = & (|\mathcal{E}_{\rm O, S}|^2 - |\mathcal{E}_{\rm X, S}|^2) \sin2\Phi_{ \rm B}
	   + 2\left[\Re(\mathcal{E}_{\rm O, S}\mathcal{E}^{*}_{\rm X, S})\cos\Delta\phi -\Im(\mathcal{E}_{\rm O, S}\mathcal{E}^{*}_{\rm X, S})\sin\Delta \phi\right]\cos 2\Phi_{\rm B} \quad , \nonumber\\[3pt]
	V & = & -2\left[\Re{(\mathcal{E}_{\rm O, S}\mathcal{E}^{*}_{\rm X, S})\sin\Delta\phi} + \Im{(\mathcal{E}_{\rm O, S}\mathcal{E}^{*}_{\rm X, S})}\cos\Delta\phi \right]  \quad , \nonumber
 \end{eqnarray}
where $\Re$ and $\Im$ respectively denote the real and imaginary components.
These equations can be applied to any altitude simply by the replacement \teq{r_{\rm rec}\to r} in the determination of \teq{\Delta \phi} and \teq{\Phi_{\rm B}}. From Equation~(\ref{eq:IQUV_rec}), it is evident that the intensity remains invariant during propagation, i.e., $dI / ds = 0$, as it should.   For highly magnetized NSs with \teq{B\gtrsim B_{\rm cr}}, $r_{\rm rec} \gg R_{\rm NS}$ (see Section~\ref{sec:recoupling}), so that the photons are propagating radially out from the star by the time they reach the recoupling surface.  Then \teq{\Phi_{\rm B}} is constant at even higher altitudes and the polarization information at infinity is essentially identical to that observed at $r = r_{\rm rec}$.  This circumstance does not arise for weakly-magnetized NSs, since for them \teq{r_{\rm rec}} is of the order of a few stellar radii, and polarization evolution outside the recoupling radius is in fact influenced by curved spacetime.

Assuming that $d \Phi_{\rm B}/ds  \approx 0$ at altitudes $r < r_{\rm rec}$, then taking the derivatives of $(Q, U, V)$ with respect to $s$ results in
\begin{equation}
   \frac{dQ}{ds} \; =  \; - \frac{\omega \Delta n}{c} V\sin 2\Phi_{\rm B}  \quad  , \  \ 
   \frac{dU}{ds} \; = \;  \frac{\omega  \Delta n}{c} V \cos 2\Phi_B  \quad , \  \
   \frac{dV}{ds} \; = \; \frac{\omega  \Delta n}{c} \Bigl(Q\sin 2\Phi_{\rm B} - U\cos 2 \Phi _B\Bigr) \quad .
 \label{eq:dQUV_1}	
\end{equation}
In delivering these forms, the only \teq{s} derivatives enter through the complex phase factors in Equation~(\ref{eq:EX_rc}), generating the simple \teq{(\omega /c) \Delta n} factors. We note the extra minus sign on the right-hand side of the evolution equations above compared to Equations (81)-(83) in \citet{Adelsberg-2006-MNRAS}; this sign difference arises from the opposite sign convention that we use for the Stokes parameter $V$.

The second, alternative approach is to follow the evolution of the Stokes parameters during propagation by considering the rotation of the polarization vector ${\bf S} = (Q, U, V)$ in the Poincare sphere \citep{Kubo1981, Kubo1983, Kubo1985}. For this geometrical optics protocol, the relation between the birefringence vector $\bf{\Omega}_{\rm B}$ and the polarization vector ${\bf S} = (Q, U, V)$ is given by Equation~(2) in \cite{Heyl2002}:
\begin{equation}
	\frac{d\textbf{S}}{ds} \; = \; \boldsymbol{\Omega}_{\rm B} \times {\bf{S}} 
	\quad , \quad
	\boldsymbol{\Omega}_{\rm B} \; = \; 
	       - \frac{\omega \Delta n}{c}\Bigl( \cos 2 \Phi_{\rm B}, \sin 2 \Phi_{\rm B}, 0\Bigr)
	 \quad .
 \label{eq:dS}
\end{equation}
The birefringence vector $\boldsymbol{\Omega}_{\rm B}$ (which has the dimensions of inverse length) was derived by \citet{Heyl-2000-MNRAS} based on the dielectric and permeability tensors of a magnetized vacuum in the formalism of \citet{Kubo1983}, and it is proportional to the \teq{\Delta n} that is given in Equation~(\ref{eq:deltan}).   One can easily show that Equation~(\ref{eq:dS}) is equivalent to Equation~(\ref{eq:dQUV_1}), which was derived from the evolution of the mode amplitudes.  It is therefore evident that the difference in the refractive indices of the O and X modes causes the polarization vector $\bf{S}$ to precess rapidly about the vacuum birefringence vector at a ``rate'' of $|\boldsymbol{\Omega}_{\rm B}|$.

\subsection{The recoupling radius}
 \label{sec:recoupling}
During photon propagation through the magnetosphere, as the magnetic field and photon directions change, $\boldsymbol{\Omega}_{\rm B}$ rotates, and the Stokes vector $\bf{S}$ tracks the birefringence vector adiabatically. In such an adiabatic regime, the two polarization modes are decoupled, as described above. At higher altitudes, the vacuum birefringence weakens.  Eventually, when the photon is sufficiently far from the NS, $|\boldsymbol{\Omega}_{\rm B}|$ drops below \teq{\vert d\Phi_{\rm B}/ds\vert} and Equation~(\ref{eq:dS}) must be replaced by the more general polarization evolution form in Equation~(\ref{eq:dE_ds}).  Then, $\bf{S}$ will no longer be able to follow $\boldsymbol{\Omega}_{\rm B}$, and the modes begin to recouple; see \cite{Heyl-2000-MNRAS, Heyl2002} for details. The adiabatic evolution for \teq{r < r_{\rm rec}} is valid if
\begin{equation}
	|\boldsymbol{\Omega}_{\rm B} | \; \gg \;\frac{1}{|\boldsymbol{\Omega}_{\rm B} |} \left| \frac{\partial |\boldsymbol{\Omega}_{\rm B} |}{\partial s} \right| \quad . \label{eq:adiabatic_HS}
\end{equation}
\citet{Heyl-2000-MNRAS, Heyl2002} formally define the recoupling radius (termed the polarization-limiting radius) as the solution of the identity
\begin{equation}
	\bigl|\boldsymbol{\Omega}_{\rm B}(r = r_{\rm rec})\bigr| 
    \; = \;  \frac{1}{2}\frac{1}{|\boldsymbol{\Omega}_{\rm B}(r = r_{\rm rec})|} \left| \left .\frac{\partial |\boldsymbol{\Omega}_{\rm B} |}{\partial s} \right|_{r = r_{\rm rec}} \right| \quad .
 \label{eq:recoupling_1}
\end{equation}
In the weak field limit, \teq{B\lesssim B_{\rm cr}}, which is clearly appropriate at sufficiently large distances from the NS surface, Equation~(\ref{eq:deltan}) indicates that \teq{\Delta n \propto B^2}.  Then, assuming that the recoupling happens at $r = r_{\rm rec} \gg R_{\rm NS}$, the photon travels radially, and $\theta_{\rm kB}$ is fixed along the high altitude portion of its trajectory.  Therefore, Equation~(\ref{eq:recoupling_1}) becomes
\begin{equation}
	\bigl|\boldsymbol{\Omega}_{\rm B}(r = r_{\rm rec})\bigr| 
    \; = \; \frac{1}{B(r = r_{\rm rec})} \left |\left. \frac{\partial B}{\partial r} \right|_{r = r_{\rm rec}}  \right| \quad .
 \label{eq:recoupling_2}
\end{equation}
In the domain where $\Psi = \rSch / r \ll 1$, the GR form of ${\boldsymbol{B}}$ in Equation~(\ref{eq:BGR}) reduces to the familiar flat spacetime dipolar form.  For \teq{\mu_{\rm B} = B_p\rns^3/2} being the magnetic dipole moment of the NS, a detailed derivation in Appendix~\ref{subsec:r_rec_HS} shows that Equation~(\ref{eq:recoupling_2}) leads to the following expressions for the recoupling radius
\begin{equation}
	r_{\rm rec} \; \approx \; \biggl( \frac{\alpha_{\rm f}}{45} \, \frac{\nu}{c} \biggr)^{1/5}\left(\frac{\mu_{\rm B} \sin\theta}{B_{\rm cr}}\right)^{2/5} 
    \; \approx \; 80 \,\rns \,\biggl( \dover{h\nu }{2\, \hbox{keV}}  \biggr)^{1/5} 
    \biggl( \dover{B_p\, \sin\theta}{10^{14}\, \hbox{G}} \biggr)^{2/5}
 \label{eq:r_rc2}
\end{equation}
for \teq{\rns = 10^6}cm.  Herein, \teq{\nu =\omega / (2\pi)} is the photon frequency, and \teq{\theta} is the magnetic colatitude. The first expression of $r_{\rm rec}$ in Equation~(\ref{eq:r_rc2}) is identical to that in Equation~(4) in \citet{Heyl2002}.  The second form indicates that 
\teq{r_{\rm rec}\sim 20-100 \rns} for 2 keV X rays (\teq{\nu \approx 5 \times 10^{17}}Hz) at all colatitudes outside magnetars except very near the magnetic axis. Specifically, since it is derived with the \teq{B(r = r_{\rm rec}) \lesssim B_{\rm cr} } form for \teq{\Delta n}, it applies for \teq{r_{\rm rec} \gtrsim 2-3\rns} and therefore for colatitudes \teq{\theta \gtrsim 10^{-2}} degrees in magnetars with \teq{B_p \gtrsim 10^{14}}Gauss and a dipolar field morphology.  Accordingly, very near the magnetic axis (i.e., over the pole), ideally one should employ the \teq{B \gg B_{\rm cr}} form for the birefringence \teq{\Delta n} in determining \teq{r_{\rm rec}}.  Yet, in practice, when integrating over large patches of the stellar surface in forming soft X-ray emission signatures for magnetars, the contribution from so near the pole is minuscule, rendering such a refinement unnecessary.   

In a separate study, \cite{Adelsberg-2006-MNRAS} employed a different criterion to define the recoupling radius. 
In their approach, the adiabatic condition and the recoupling (polarization limiting) radius are respectively expressed via
\begin{equation}
     \bigl| {\boldsymbol{\Omega}_{\rm B}} \bigr| \; \gg \;  2 \frac{d\Phi_{\rm B}}{ds}
     \quad \hbox{and}\quad
	\bigl| \boldsymbol{\Omega}_{\rm B}(r = r_{\rm rec}) \bigr| 
    \; =\; 2 \left. \frac{d\Phi_{\rm B}}{dr}\right|_{r = r_{\rm rec}}  \quad .
 \label{eq:rpl_AL0}
\end{equation}
This scheme essentially compares the mode-mixing off-diagonal terms in Equation~(\ref{eq:dE_ds}) with the diagonal ones, using the definition of the birefringence vector in Equation~(\ref{eq:dS}).  Since, at high altitudes, the field direction changes appreciably with time, this criterion naturally introduces a dependence on the spin period of the star.  In their paper, \cite{Adelsberg-2006-MNRAS} also considered the regions far away from the NS surface that the photon travels radially. Therefore,  $\Phi_{\rm B}$ identifies with the angle between $\boldsymbol{B}$ at any rotational phase and the plane made by the x-z plane defined in Equation~(\ref{eq:coordinate}), so that
\begin{equation}
	\tan \Phi_{\rm B} \; = \; \frac{\boldsymbol{B} \cdot \hat{y}}{\boldsymbol{B} \cdot \hat{x}} \; = \;  \frac{\sin \alpha \sin \Phi}{- \cos \alpha \sin \zeta + \cos \Phi \cos \zeta \sin \alpha } \quad , \label{eq:tanphiB_2}
\end{equation}
where $\alpha$ is the inclination angle of the magnetic dipole moment $\hat{\boldsymbol{\mu}}_B$ to the spin axis $\hat{\boldsymbol{\Omega}}$, and $\zeta$ is the angle between the line of sight  $\hat{\boldsymbol{k}}_{\infty}$ and $\hat{\boldsymbol{\Omega}}$.  Also, $\Phi = \Omega t$ denotes the rotational phase, with $\Omega = 2\pi/P$ being the angular frequency of the NS's rotation, and \teq{P} its spin period.  We adopt the convention where $t = 0$ corresponds to the moment when $\hat{\boldsymbol{\mu}}_B$, $\hat{\boldsymbol{\Omega}}$, and $\hat{\boldsymbol{k}}_{\infty}$ lie on the same plane, termed the meridional plane.  Substituting Equation~(\ref{eq:tanphiB_2}) into Equation~(\ref{eq:rpl_AL0}) gives
\begin{equation}
     r_{\rm rec} \; \approx \; 151.5 \left(\frac{\edithere{P_1}E_1 B^2_{14}F_B}{ F_{\varphi}}\right)^{1/6} R_{\rm NS}  \quad  ,
 \label{eq:rplAP_final}
\end{equation}
where $E_1 = E_{\gamma} /(1\mathrm{ keV}) $, with $E_{\gamma}$ being the photon energy, $B_{14} = {B_p}/{(10^{14} G)}$, \edithere{and $P_1 = P/(1s)$.}  Also, $F_{B}$ and $F_{\phi}$ are dimensionless functions of geometry angles.  The detailed derivation of Equation~(\ref{eq:r_rc2}) together with the expressions of $F_{B}$ and $F_{\phi}$ can be found in Appendix~\ref{sec:appendix}. 

Equations~(\ref{eq:adiabatic_HS}) and (\ref{eq:rplAP_final}) provide complementary estimates of the recoupling radius.  Comparing them, one quickly discerns differences on the right-hand side of these equations.  Specifically, \citet{Heyl-2000-MNRAS, Heyl2002} defined the adiabatic condition by comparing the magnitude of the vacuum birefringence vector, $|\boldsymbol{\Omega}_{\rm B}|$, with its rate of change (or equivalently, the rate of change of the magnetic field strength). Thus, the corresponding recoupling radius, $r_{\rm rec}$, as described in Equation~(\ref{eq:r_rc2}),  depends on the distance from the NS and the direction of photon propagation relative to the magnetic dipole axis. On the other hand, \cite{Adelsberg-2006-MNRAS} compared $|\boldsymbol{\Omega}_{\rm B}|$ with the change in the magnetic field direction due to the rotation of the NS. As a result, $r_{\rm rec}$ in Equation~(\ref{eq:rplAP_final}) also has a dependence on the NS rotation frequency. Generally, the recoupling radii obtained using these two approaches are of the same order of magnitude.  Yet for slow rotators like magnetars,  the prescription by \citet{Heyl-2000-MNRAS, Heyl2002} typically yields a smaller recoupling radius compared to that by \citet{Adelsberg-2006-MNRAS}.
Accordingly, in this work, we adopt the prescription by \cite{Heyl-2000-MNRAS, Heyl2002} for the recoupling radius, as encapsulated in Equation~(\ref{eq:r_rc2}), since the photon polarizations recouple below altitudes where stellar rotation significantly influences the character of the birefringence.

We remark that \citet{Heyl-2000-MNRAS, Heyl2002} and \cite{Adelsberg-2006-MNRAS} both assumed an abrupt transition from the decoupled state to the recoupled configuration for the two polarization modes. Indeed, the factors of $1/2$ and $2$ on the right-hand side in Equations~(\ref{eq:recoupling_1}) and (\ref{eq:rpl_AL0}), respectively, are somewhat subjective choices, and they introduce a sharp boundary beyond which the adiabatic evolution is no longer valid. In reality, we would expect the transition to occur more gradually \citep{Fernandez-2011-ApJ}. Another limitation is that the expressions of the recoupling radius in both approaches were derived without accounting for GR effects. This approximation is valid only for highly magnetized NSs, where the recoupling occurs far from the surface that GR effects are negligible. However, this is less accurate for NSs with lower magnetizations. A precise description of the recoupling radius requires following each photon in its curved trajectory in the Schwarzschild metric and solving the evolution equations numerically. This is beyond 
the scope of this paper. 

\begin{figure}[t]
	\centering
	\includegraphics[width=1.0\linewidth]{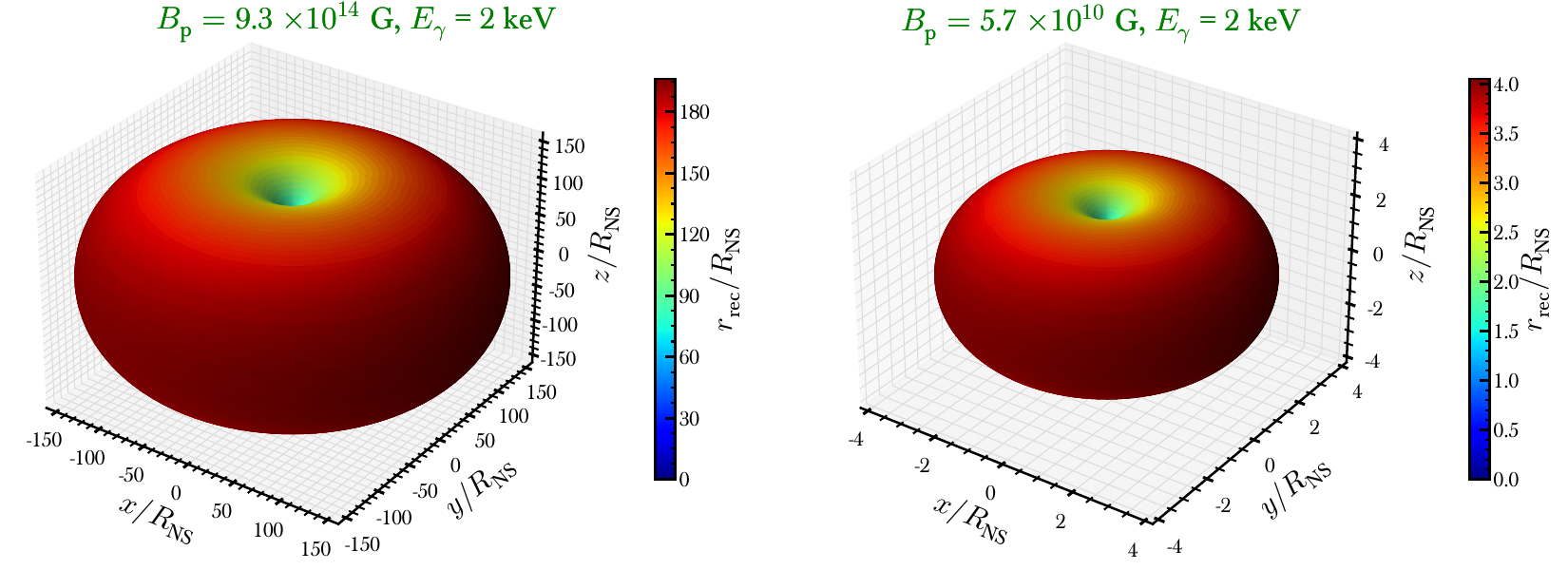}
	\caption{Recoupling surfaces for photons with an energy of $E_{\gamma} = 2$ keV emitted from magnetar 1RXS J1708-4009 with a polar magnetic field strength of $B_p = 9.3 \times 10^{14}$ G (left panel) and CCO RX~J0822.0-4300 with $B_p = 5.7 \times 10^{10}$ G (right panel). In both panels, the NS is positioned at (0, 0, 0). The colorbar on the right of each panel represents the recoupling radius in units of the NS radius, $r_{\rm rec} / R_{\rm NS}$. The z-axis in both panels align with the magnetic dipole moment. We employed the prescription by \cite{Heyl-2000-MNRAS} as in Equation~(\ref{eq:r_rc2}). }
	\label{fig:fig-recoupling}
\end{figure}

In Figure~\ref{fig:fig-recoupling}, we display the recoupling surfaces obtained using Equation~(\ref{eq:r_rc2}) for photons with an energy of $E_{\gamma} = 2$~keV for a polar magnetic field strength of $B_p = 9.3 \times 10^{14}$~G (left panel) and $B_p = 5.7 \times 10^{10}$~G (right panel).  These values respectively correspond to the polar magnetic field strengths measured by an observer at infinity for the magnetar 1RXS J1708-4009 and the CCO pulsar RX~J0822.0-4300, whose intensity and polarization characteristics will be discussed in the following section. The colorbar on the right of each panel represents the recoupling radius in units of the NS radius, $r_{\rm rec} / R_{\rm NS}$.  In both panels, the NS is positioned at the origin, and the z-axis aligns with the instantaneous magnetic dipole moment. Since $r_{\rm rec}$ only varies with magnetic colatitude $\theta$, the recoupling surface is axi-symmetric  around the magnetic dipole moment. As the vacuum birefringence effect is zero for photons traveling along the magnetic field, for photons traveling along the dipole moment axis, $r_{\rm rec} = 0$.  This marks the appearance of a ``toroidal hole'' in the illustrated surfaces along the instantaneous dipole moment vector.
Furthermore, except for a small solid angle surrounding
$\hat{\boldsymbol{\mu}}_{\rm B}$, in the magnetar case (left panel), $r_{\rm rec}$ is typically larger than $100R_{\rm NS}$. In this case, we can expect vacuum birefringence in the magnetosphere to preserve the surface polarization signature. On the contrary, for the lower magnetization case depicted on the right, the recoupling occurs at modest or small alitudes, and therefore, we expect a less significant impact of vacuum birefringence on the emergent polarization characteristics. The values we found for $r_{\rm rec}$ are general agreement with those obtained in \citet{Taverna2015}: see their Figure~1.   We note that their presentation did not account for the dependence of the recoupling radius on the relative direction (i.e., \teq{\theta}) between the photon and the magnetic field.

\section{Results and Discussion}
 \label{sec:result-and-discussion}

The surface of a NS can be divided into a grid of small or infinitesimal patches corresponding to different magnetic longitudes and colatitudes, where the emission from each patch can be treated using the slab geometry formalism described in Section~\ref{sec:radiative-transfer}. Incorporating the elements outlined in Sections~\ref{sec:radiative-transfer} and \ref{sec:propagation},  our  {\sl MAGTHOMSCATT} simulation diligently tracks the complex electric field vector of each photon during its scattering process in the atmosphere as well as propagation in the magnetosphere, where both GR and QED effects are taken into account. We modeled the intensity and polarization signals observed in a particular direction at infinity by integrating over different locales within the emission regions. The flexibility of our Monte Carlo simulation in treating the magnetic field direction allows for modeling various hot spot configurations, including different shapes and locations. This aspect will be explored in future work that focuses on modeling specific sources. Here, \edithere{we illustrate a range of results and our fitting protocols.  We restrict our surface emission geometry} to two antipodal polar caps, where the regions from each magnetic pole to colatitude of $\theta_{\rm cap}$ are assumed to be hotter than the rest of the star, with their emissivities uniform across these caps.

The Stokes parameters ($I, Q, U, V$) at infinity are calculated in the coordinates defined in Equation~(\ref{eq:coordinate}), from which the linear PD, $\Pi_l = \sqrt{Q^2 / I^2 + U^2 / I^2}$, and PA, $\chi = (1/2)\, \arctan{(U/Q)}$, can also be calculated. In this exposition, the magnetic inclination angle $\alpha$ ranges from $0^{\circ}$ to $90^{\circ}$ in increments of $5^{\circ}$.  Meanwhile, the viewing angle $\zeta$ to the spin axis \teq{\boldsymbol{\Omega}} varies from $0^{\circ}$ (``north'' pole view) to $180^{\circ}$ (``south'' pole view), in increments of $4^{\circ}$. The rotational phase, $\Phi \in [0, 360^{\circ}]$ for each cycle, increases in steps of $4^{\circ}$ as well.
The number of photons collected at infinity, $\mathcal{N}_{\rm rec}$, is chosen to be sufficiently high such that simulating larger counts does not statistically affect the results.

In Section~\ref{sec:intensity}, we present our predictions for the intensity pulse profiles using different geometric parameters.  By comparing our predictions with the phase-resolved intensity data for magnetar 1RXS  J1708-4009, we provide constraints on $\alpha$, $\zeta$, and $\theta_{\rm cap}$.  This exposition is intended as a proof of principal, and demonstrates the power of intensity pulse profile modeling; yet it is not designed as a detailed study of this source, which can be found in Stewart et al (in prep.).  In Section~\ref{sec:polarization}, we discuss the influence of vacuum birefringence within the magnetosphere on observable polarization characteristics, focusing on magnetar 1RXS J1708-4009 and CCO pulsar RX~J0822.0-4300.

\subsection{Modeling intensity pulse profiles}
\label{sec:intensity}

\begin{figure}[ht]
	\centering
	\includegraphics[width=1.0\linewidth]{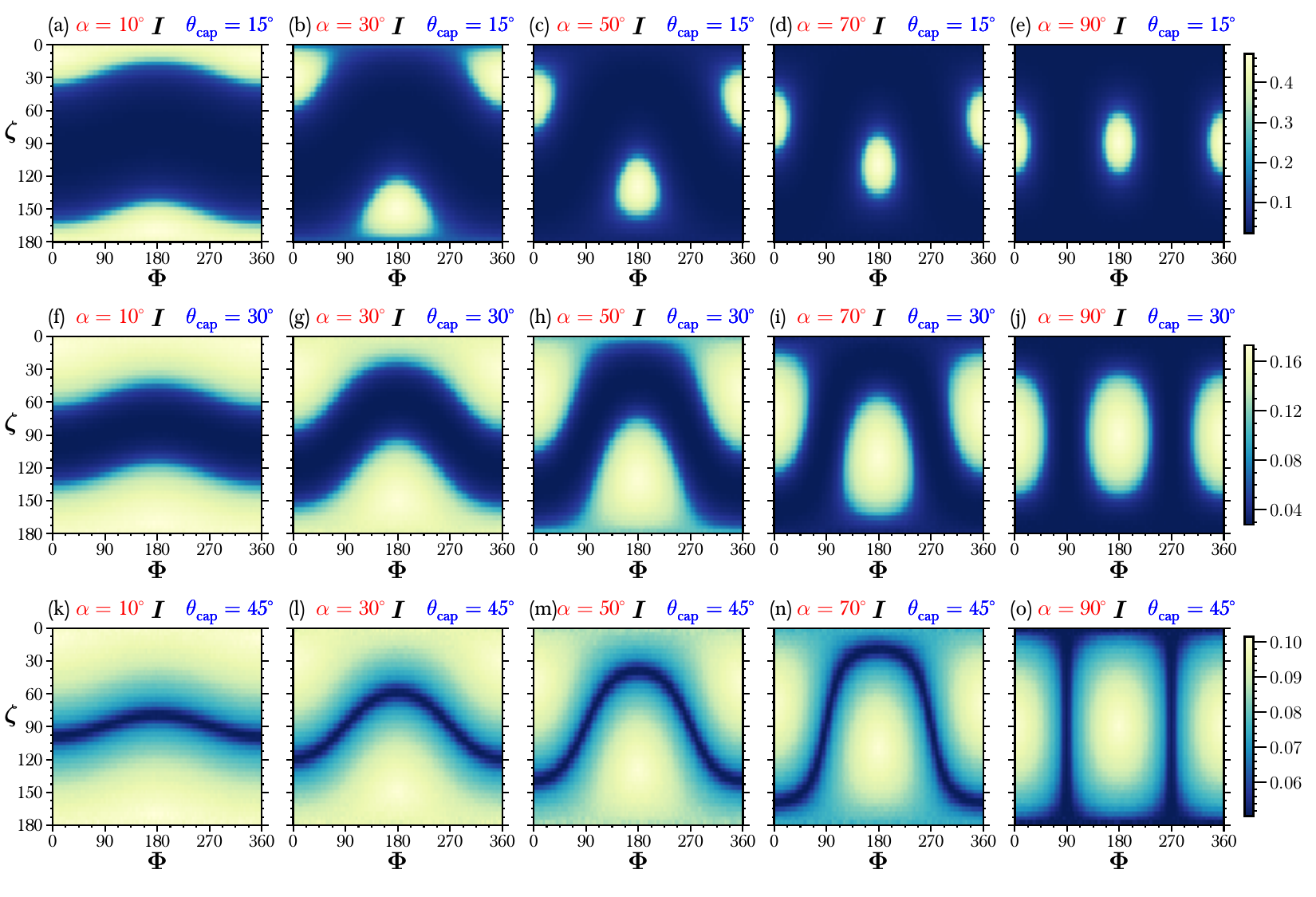}
	\caption{Intensity sky maps obtained from the modeling of two uniform antipodal polar caps extending from the respective magnetic poles to colatitudes $\theta_{\rm cap} = 15^{\circ}$ (top row), $\theta_{\rm cap} = 30^{\circ}$ (middle row), and $45^{\circ}$ (bottom row), as functions of viewing angle $\zeta$ and rotational phase $\Phi = \Omega t$.  The five columns correspond to five different values of the inclination angle, $\alpha =10^{\circ}, 30^{\circ}, 50^{\circ}, 70^{\circ}, 90^{\circ}$, as labelled.  To approximate the magnetar domain, the simulations were performed at $\omega /\wcyc = 0.01$ in the LIF at the magnetic poles (see text).  The number of recorded photons  is $\mathcal{N}_{\rm rec} = 10^8$. }
	\label{fig:intensity-skymap-magnetar}
\end{figure}

In Figure~\ref{fig:intensity-skymap-magnetar}, we display the intensity sky maps, i.e., relative emission probabilities, as functions of the viewing angle, $\zeta$, on the y-axis, and the rotational phase, $\Phi = \Omega t$, on the x-axis.  These were obtained for $\mathcal{N}_{\rm rec}= 10^8$ photons, from models of two uniform antipodal polar caps extending from the respective magnetic poles to colatitudes  $\theta_{\rm cap}$ = 15$^{\circ}$  (top row), $\theta_{\rm cap}$ = 30$^{\circ}$ (middle row), and 45$^{\circ}$ (bottom row). 
Five columns correspond to five selected values of the inclination angle, $\alpha =10^{\circ}, 30^{\circ}, 50^{\circ}, 70^{\circ}, 90^{\circ}$, left to right. The intensity $I$ is normalized such that the integral over the solid angle equals one: 
\begin{equation}
    \int_{0}^{2\pi} \int_0^{\pi} 
    I \sin{\zeta} \, d\zeta \, d\Phi 
    \; =\; 1 \quad .
 \label{eq:intensitiy_norm}
\end{equation}
The variation in the intensity is noticeably reduced when increasing the capsize from $15^{\circ}$ to $45^{\circ}$, as is evident from the comparison of the top, middle, and bottom row panels.  Additionally, the intensity peaks when the observer's line of sight aligns with the magnetic dipole moment axis, as previously  discussed in \cite{Hu-2022-ApJ}. This alignment corresponds to a viewing angle of $\zeta = \alpha$ for $\Phi = 0^{\circ}$ and $\Phi = 360^{\circ}$, and $\zeta = 180^{\circ} - \alpha$ for $\Phi =180^{\circ}$.  For large inclination angles, $\alpha > 30^{\circ}$, the intensity exhibits strong pulsations, resulting in pronounced peaks.  Furthermore, when $\alpha \sim 90^{\circ}$, we can observe two peaks within a single cycle, since both polar caps are equally visible overall, yet at complementary phases.  On the other hand, for small inclination angles, $\alpha \lesssim 30^{\circ}$, the intensities are more spread out, and there is only one peak per cycle. Thus, based on these sky maps and the characteristics of the observed intensity pulse profiles, namely pulse fraction and number of peaks per cycle, one can navigate the appropriate values of $\theta_{\rm cap}$, $\alpha$, and $\zeta$.

To accurately determine the model that best matches the data, we performed a reduced $\chi^2$ statistical fit.  The widely-used $\chi^2$-function per degree of freedom \citep{Arfken-Weber-2005,Dobaczewski2014} is employed throughout this paper.  This function has three key model parameters ($\theta_{\rm cap}$, $\alpha$, $\zeta$) that can be varied, and can be defined along with its associated probability function \teq{P(\theta_{\rm cap}, \alpha, \zeta)} as follows:
\begin{equation}
	\chi_{I}^2 \bigl(\theta_{\rm cap}, \alpha, \zeta \bigr) 
    \; = \; \frac{1}{N_{\rm dof}}\sum_{j =1}^{N_{\rm d}}\frac{\bigl[I_j(\theta_{\rm cap}, \alpha, \zeta) - I^{\rm obs}_j \bigr]^2}{\sigma^2_j} 
    \quad ,\quad
    P(\theta_{\rm cap}, \alpha, \zeta) 
    \; \propto\;  \exp{\left(-\frac{\chi_{I}^2}{2} \right)} \quad .
	\label{eq:chi2} 
\end{equation}
Herein, $I_j(\theta_{\rm cap}, \alpha, \zeta)$ and $I^{\rm obs}_j$ respectively denote the simulated and observed intensity, $\sigma_j$ represents the standard deviations of the observational uncertainties, $N_{\rm d}$ is the number of data points, and $N_{\rm dof} = N_{\rm d} - N_{\rm par}$ is the number of degrees of freedom, with $N_{\rm par} = 3$ being the number of model parameters (i.e., the arguments of $I_j$).  In this fitting protocol, a Gaussian normal distribution is presumed for the data uncertainties, and the theoretical intensity is scaled by a constant such that it has the same phase-averaged normalization as that of the data. $P(\theta_{\rm cap}, \alpha, \zeta)$ will be normalized so that summing over the parameter selections (see below) yields unity, so that \teq{P} describes the viability probability for theoretical models.

Following the protocol described above, we fit our {\sl MAGTHOMSCATT} simulated light curves to intensity data for magnetar 1RXS J1708-4009. \edithere{As mentioned above, the light-curve fitting presented here is intended primarily to demonstrate the proof of concept, rather than to achieve the most refined fit. Therefore, we only focus on hot-spot configurations consisting of two antipodal and azimuthally symmetric polar caps for simplicity. For this exercise, we consider nine $\theta_{\rm cap}$ values, ranging from $5^{\circ}$ to $45^{\circ}$, in increment of $5^{\circ}$. The magnetar 1RXS J1708-4009} has a polar field strength of \teq{B_p = 9.3 \times 10^{14}}Gauss as inferred at infinity \citep[double the equatorial field identified in][]{OK-2014-ApJS}; in the LIF at the surface it has the higher value of \teq{B_p \approx 1.4 \times 10^{15}}Gauss  for our chosen compactness \teq{\Psis = 2G\mns /(c^2\rns )=0.425}.  The light curve data was obtained by XMM-Newton in the energy range of $0.5-3$ keV (Stewart et al., in prep.) with approximately 250,000 counts in total, presented here using 28 pulse phase bins. The energy-dependent root-mean-squared pulse fraction in this band remains largely stable at around 27\%.  While the pulse profile morphology exhibits slight variations with energy near the dip between $0.5-3$~keV, it displays a strong energy dependence above \teq{3} keV. The nature of this energy sensitivity is briefly discussed below, and is more fully explored in Stewart et al. (in prep.).

\begin{figure}[t]
	\centering
	\includegraphics[width=1.0\linewidth]{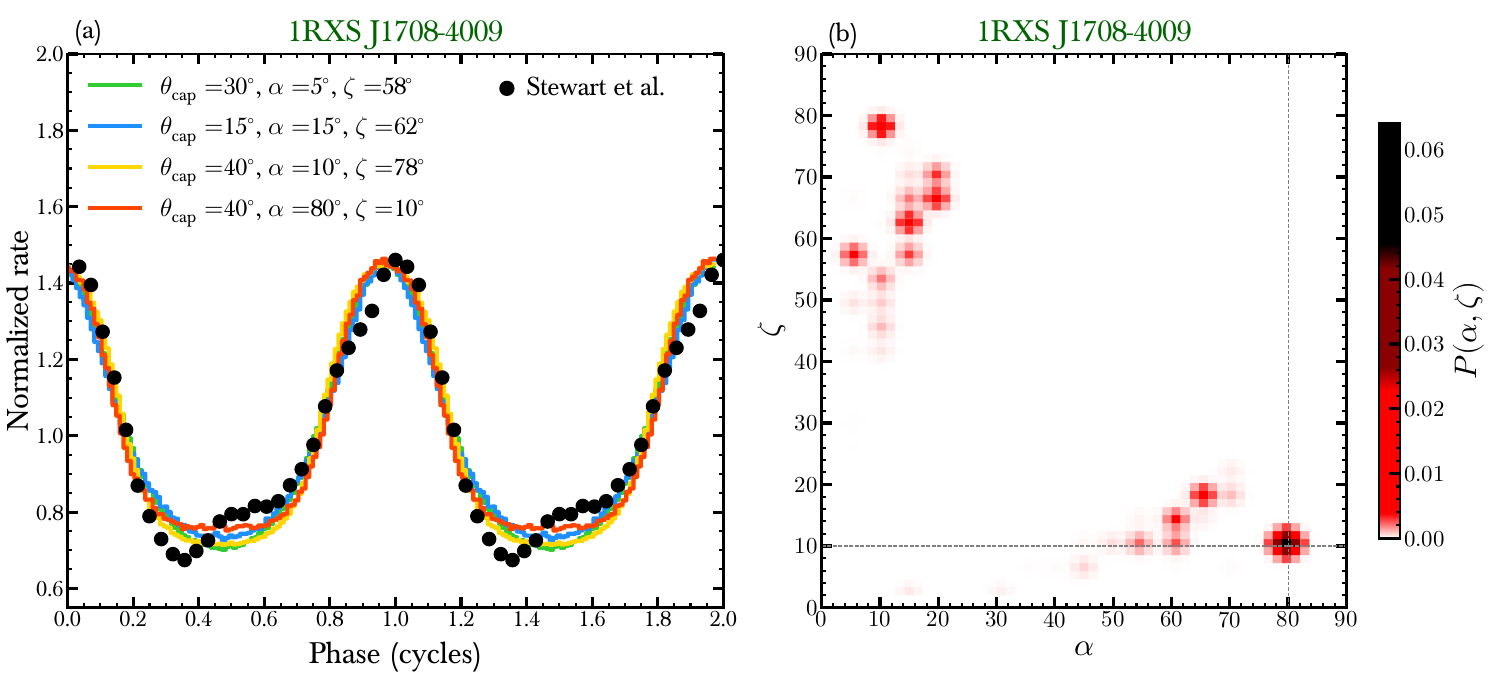}
	\includegraphics[width = 1.0\linewidth]{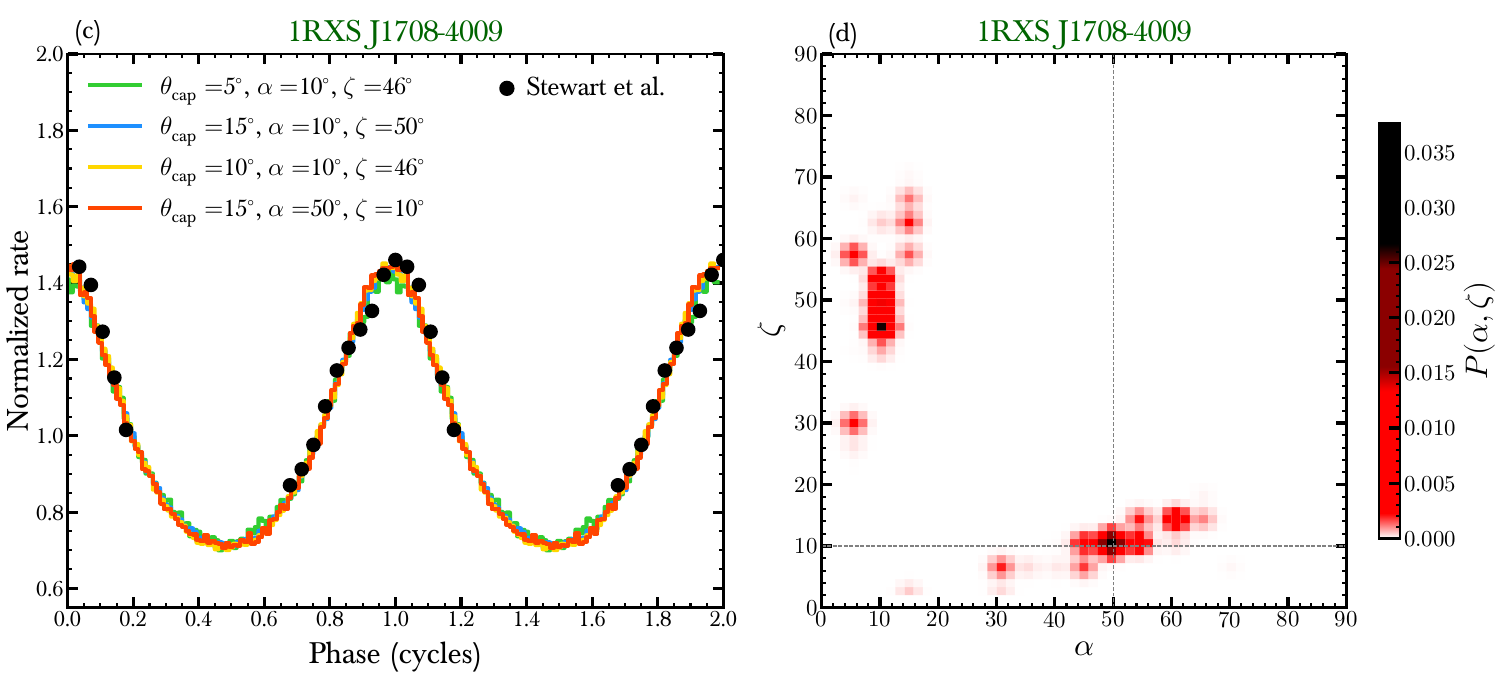}
    \caption{\edithere{Panel (a)}: Simulated pulse profiles for intensity $I$ for two antipodal polar caps extending from the respective magnetic poles to colatitudes of  $\theta_{\rm cap} = 15^{\circ}, 30^{\circ}, \edithere{40^{\circ}}$ as a function of rotational phase in units of cycles $\Phi / (2\pi)$. The black scattered points represent the data extracted from XMM in the
	energy band $0.5-3$ keV by Stewart et al. (in prep.) for   magnetar 1RXS J1708-4009. The red curve represents the parameter set ($\theta_{\rm cap}$, $\alpha$, $\zeta$) = \edithere{($40^{\circ}$, $80^{\circ}$, $10^{\circ}$)} that provides the best fit to the data. The other three curves correspond to parameter sets that also fit the data well. \edithere{Panel (b)}: Two-dimensional distribution of the geometric parameters ($\alpha, \zeta$), constrained with the observed intensity data on the left. The colorbar indicates the normalized probability density $P(\alpha, \zeta) = \mathcal{N}\sum_{i}P(\theta^i_{\rm cap},\alpha, \zeta)$, with $\mathcal{N}$ being a normalization constant such that the volume below the distribution
    equals one, i.e., $\sum P(\alpha, \zeta)\Delta\alpha \Delta\zeta = 1$. The vertical and horizontal dashed lines indicates the most probable value of $\alpha$ and $\zeta$, respectively. \edithere{Panels (c) and (d): a counterpart to panels (a) and (b), but obtained with the excised dataset, that is, omitting the intensity data points between phases $0.20-0.65$.  Note the different best fit parameters.  See text for details.}}
	\label{fig:best-intensity-fit-J1708}
\end{figure}

In Figure~\ref{fig:best-intensity-fit-J1708}, \edithere{panel (a)}, we depict our best reduced $\chi^2$-fit light curve (red histogram) alongside the normalized rate data (black points) as a function of rotational phase in units of cycle, $\Phi / (2 \pi)$. The data quality is excellent, with relative errors of less than $\sim 1\%$, which effectively makes the error bars invisible in the plot.  The best-fit case corresponds to \edithere{\teq{\alpha = 80^{\circ}} and \teq{\zeta = 10^{\circ}} for a polar cap size of $\theta_{\rm cap} = 40^{\circ}$,} the red histogram in \edithere{panel (a)}. The other three solid curves represent models that yield comparably good fits, \edithere{indicating a broad range of likely key geometrical parameters.}  The overall agreement between our simulated pulse profiles with the data is satisfactory, albeit poorer in the \teq{0.2 - 0.4} phase interval where an intensity dip is apparent.  Due to the azimuthal uniformity inherent in our identical antipodal cap assumption, the peaks and valleys in the model light curves are symmetric.  As a consequence, they cannot capture the distortion associated {with the dip, which indicates possible non-uniformities of the surface that produce the observed lightcurves.  Specifically, surface emission that is uniform in magnetic longitude generates light curves that display a \teq{\Phi/(2\pi ) \leftrightarrow 1 - \Phi/(2 \pi )} phase reflection symmetry, as is evident in Figure~\ref{fig:intensity-skymap-magnetar}; see also sky maps for various ranges of emission colatitudes in \cite{Hu-2022-ApJ,Dinh-2025-ApJ}.  One therefore deduces that the dip is perhaps caused by a lower average emissivity at a restricted range of magnetic longitudes that is centered around \teq{120^{\circ}} from the meridian.}

An alternative interpretation of the light curve structure in the \edithere{\teq{0.2-0.65}} phase range is that there is an enhancement or bump between 0.4 and \edithere{0.65} phases.  The average behavior would then have a deeper trough with a higher pulse fraction, with a higher emission signal superposed at constrained magnetic longitudes corresponding to this bump.  We performed a fit to the light curve formed by excising the \teq{\edithere{0.2}-0.65} phases using {\sl MAGTHOMSCATT} \edithere{(bottom row, Figure~\ref{fig:best-intensity-fit-J1708})}, and found \edithere{a reduction by a factor of three} in \teq{\chi_{I}^2} (because the light curve data are now more symmetric)\edithere{, particularly from \teq{\chi_{I}^2 \approx 32} to \teq{\chi_{I}^2 \approx 11} for the best fit, with \teq{\chi_{I}^2} defined in Equation~(\ref{eq:chi2}). We note that the best $\chi^2_I$ value is still significant even when fitting to the excised data because of the slight asymmetry persisting in the light curve around the pulse peak and the remarkably small uncertainties in the observed data points; this asymmetry likely captures non-uniformaties in the distribution of emissivities across the surface. The} distribution of the probability density in \teq{\alpha - \zeta} space \edithere{in this case is somewhat shifted}, albeit with similar general morphology to that displayed \edithere{in panel (b)} of Figure~\ref{fig:best-intensity-fit-J1708}.  From this ``experiment,'' we conclude that with the existing light curve data, a more detailed investigation of the hot-spot shapes and geometrical placement is warranted.  Such a future project could help discriminate between an over-luminous band of magnetic longitudes and an adjacent under-luminous longitude band in the surface emission profile, both being possible signatures of non-axisymmetric twists of the magnetic field. 

\begin{figure}[t]
	\centering
	\includegraphics[width=1.0\linewidth]{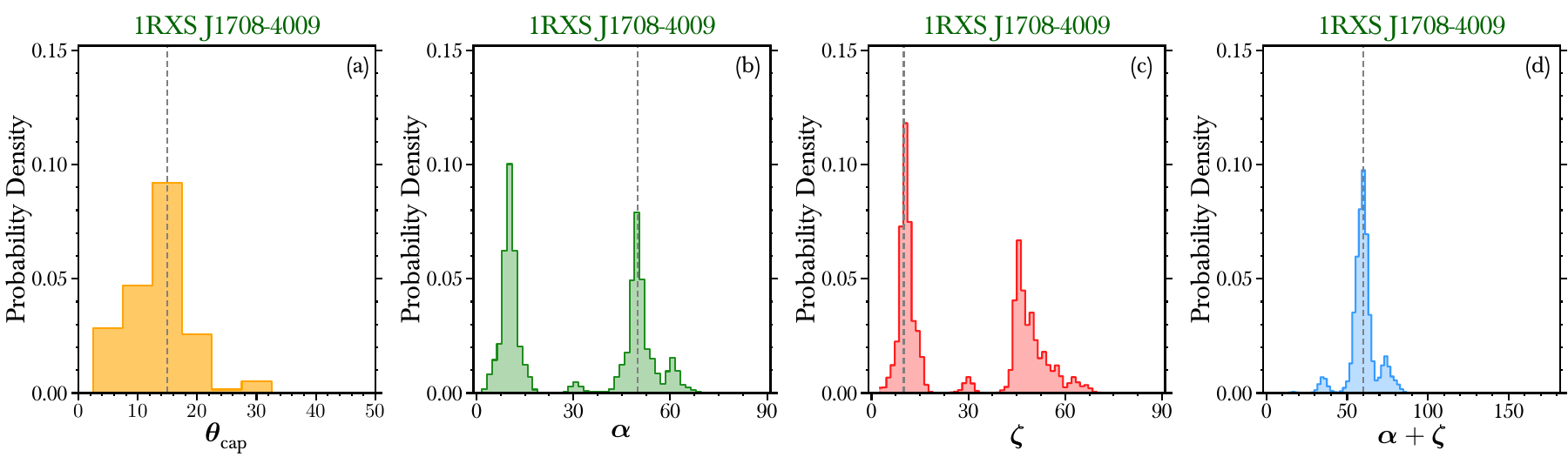}
	\caption{\edithere{Probability density distributions of the capsize $\theta_{\rm cap}$ \edithere{(panel (a))}, $\alpha$  (panel (b)),  $\zeta$  (panel (c), and $\alpha + \zeta$ (panel (d)); all distributions are normalized such that the area below each distribution equals one.   These distributions are obtained using observed intensity data for magnetar 1RXS J1708-4009 from Stewart et al. (in prep.), with the data points between phases 0.20-0.65 being excised; see text. The vertical dashed lines indicate the best-fit values.}}
	\label{fig:1d-distribution-parameter-J1708}
\end{figure}

In these illustrations, a combination of moderately small inclination angle ($\alpha \leq 20^{\circ}$) and large viewing angle ($\zeta > 50^{\circ}$), or large $\alpha$ and small $\zeta$, is preferred.  To see this more incisively, 
in \edithere{panel (b)} of Figure~\ref{fig:best-intensity-fit-J1708}, we present the two-dimensional probability density distribution of the geometric parameters, ($\alpha, \zeta$), constrained with the observed intensity data shown \edithere{in panel (a)}. The colorbar indicates the normalized probability $P(\alpha, \zeta)$, obtained by summing $P(\theta_{\rm cap}, \alpha, \zeta)$ over the 
\edithere{nine values of $\theta_{\rm cap}$.}
The probability density $P(\alpha, \zeta)$ is normalized such that the volume below the distribution equals one.   
To reduce the discreteness from finite binning and produce visually smoother distributions, we subdivided each $\alpha$ and $\zeta$ bin by a factor of three and applied Gaussian smoothing when plotting the right panel of Figure~\ref{fig:best-intensity-fit-J1708} and  Figure~\ref{fig:1d-distribution-parameter-J1708}.  The dashed vertical and horizontal lines correspond to the values of $\alpha$ and $\zeta$ of the most probable configuration. 
A noticeable feature of this probability map is that it is not absolutely symmetric under the interchange of \teq{\alpha} and \teq{\zeta}. As discussed in \cite{Hu-2022-ApJ}, since the angle between the line of sight and the instantaneous magnetic axis is symmetric in \teq{\alpha} and \teq{\zeta}, if the surface emission is independent of magnetic longitude, all model light curves will be invariant under the interchange \teq{\alpha \leftrightarrow \zeta}. The fit probability distributions \edithere{in panel (b) of} Figure~\ref{fig:best-intensity-fit-J1708} also capture the information of the asymmetry in the observational light curve \edithere{in panel (a)}, again indicating that there is a longitudinal asymmetry in the heating of the surface. \edithere{The $\alpha-\zeta$ distribution becomes more symmetric when using the excised data, that is, without the data points between phases $0.2-0.65$, as shown in panel (d). Nevertheless, the symmetry is not perfect due to the slight asymmetry remaining near the intensity peak in the observed data.}

The individual probability distributions of $\theta_{\rm cap}$, $\alpha$, $\zeta$, and their sum ($\alpha+\zeta$), \edithere{obtained with the excised data,} are shown in Figure~\ref{fig:1d-distribution-parameter-J1708}. It is evident from the panel (a) of this Figure that $\theta_{\rm cap} = 15^{\circ}$ is favored by the intensity data\edithere{. The} negligible probability for $\theta_{\rm cap} = 45^{\circ}$ indicates that this large cap size is incompatible with the given observational data.  In our previous analysis in \cite{Baring2024b}, we found that fitting simulated intensity pulse profiles to the {\sl IXPE} pulse profile data from \cite{Zane-2023-ApJ} yielded comparable probabilities for the cases $\theta_{\rm cap} = 45^{\circ}$ and $\theta_{\rm cap} = 30^{\circ}$, with the larger capsize favoring $\alpha \sim 60^{\circ}$ and $\zeta \sim 40^{\circ}$. However, in that ensemble of simulations, the effective optical depth was fixed at  $\tau_{\rm eff} = 6$ across all magnetic colatitudes. While this approximation is appropriate for small caps, it becomes increasingly inaccurate for larger caps, where the dependence of $\tau_{\rm eff}$ on magnetic colatitude cannot be neglected \citep{Dinh-2025-ApJ}. We refined this protocol in \cite{Dinh-2025-ApJ}, where we determined the optimal values of $\tau_{\rm eff}$ for different magnetic colatitudes spanning from the magnetic pole to the equator.   This advance led to the delivery of precision surface anisotropies and emissivities for all the simulation runs produced for this paper.  As a result, the $\theta_{\rm cap} = 45^{\circ}$ scenario is no longer favored, and the corresponding geometric parameters are excluded, leading to more constrained distributions for $\theta_{\rm cap}$, $\alpha$ and $\zeta$. 

For a NS radius of \teq{R=10}km, the cap surface areas \teq{2\pi (1 - \cos \theta_{\rm cap})} are 21.4 km$^{2}$ and 84.2 km$^{2}$ for \teq{\theta_{\rm cap} = 15^{\circ}} and \teq{\theta_{\rm cap} = 30^{\circ}}, respectively.  These can be compared with the area of 51.5 km$^{2}$ deduced by Stewart et al. (in prep.) from the phase-averaged flux and spectral temperature using the Stefan-Boltzmann law and a source distance of 3.85 kpc \citep{Durant-2006-ApJ}.  While the ranges of possible inclination angles $\alpha$ (panel (b)) and  viewing angles $\zeta$ (panel (c)) are quite broad, ($\alpha +\zeta$) distribution (panel (d)) is well constrained.  It is clearly concluded that using intensity data alone, 1RXS J1708-4009 is an oblique rotator that is viewed at moderate to large angles to its spin axis.  Polarization data, where available, can be incorporated in order to further constrain these parameter distributions; this is addressed in the paper by Stewart et al. (in prep.). 

An interesting study of a 2.2 hour duration flare episode of 1RXS J1708-4009 was provided in \cite{Younes-2020-ApJ}.  Therein, the long term NuSTAR pulse profile for data above 3 keV was subtracted from the NuSTAR flare light curve yielding a symmetric, double-peaked profile.  This excess was then fit with a symmetric, antipodal hot spot model that employed a phenomenological angular emission profile on caps of around \teq{\theta_{\rm cap} \sim 20^{\circ}} extent.  \cite{Younes-2020-ApJ} concluded that the best fit solution was for \teq{\zeta = 60^{\circ}}, with the flare state hot spots offset by \teq{60^{\circ}} from the spin axis, albeit using NuSTAR data of much poorer count statistics than those of the XMM-Newton data employed here.  A similar geometry solution was obtained using {\sl MAGTHOMSCATT} in the paper of \cite{Hu-2023-IAUS363}.  If the two locales for the hot spot enhancements are interpreted as being at the magnetic poles, the baseline assumption of \cite{Younes-2020-ApJ} and \cite{Hu-2023-IAUS363}, then a magnetic inclination of \teq{\alpha \sim 60^{\circ}} would be inferred.  While this would appear to conflict with the preferred solution space in Figures~\ref{fig:best-intensity-fit-J1708} and~\ref{fig:1d-distribution-parameter-J1708}, it is apparent in Stewart et al. (in prep.) \edithere{that quiescent emission above 3 keV is mostly non-thermal: only 6.5\% of the phase-averaged flux above 3 keV derives from a blackbody spectral component.  Accordingly, emission at these higher energies is not necessarily connected} to surface atmospheres.  This poses an intriguing conundrum whose resolution requires a future dedicated study of hot spot geometries using energy-dependent pulse profiles of 1RXS J1708-4009 in both quiescent and flaring episodes.

\begin{figure}[bh]
	\centering
	\includegraphics[width=1.0\linewidth]{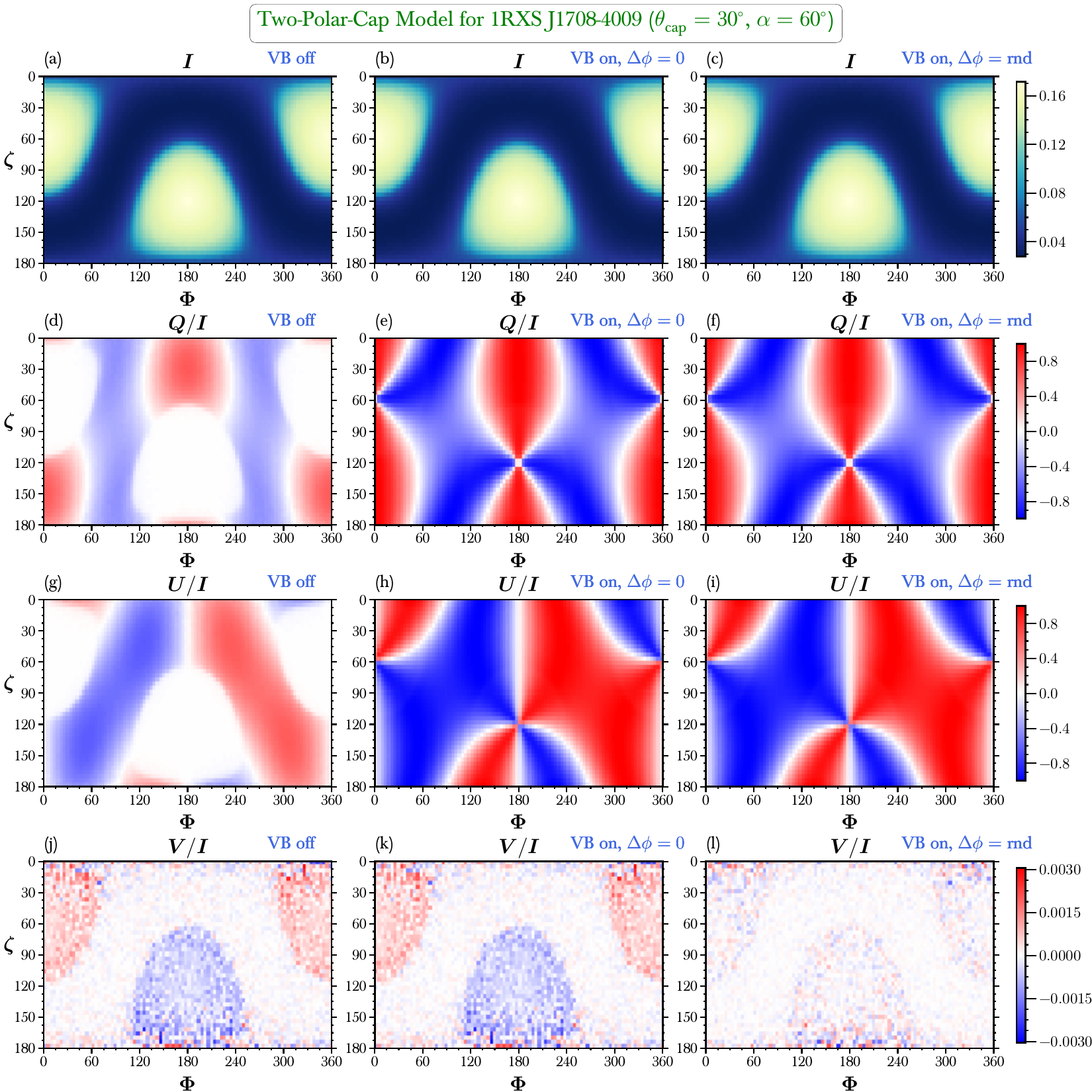}
	\caption{Intensity and polarization ($Q/I$, $U/I$, $V/I$) sky maps in \teq{\zeta} versus \teq{\Phi = \Omega t} space, obtained from the modeling of two uniform antipodal polar caps extending from the respective magnetic poles to colatitudes $\theta_{\rm cap} = 30^{\circ}$ for magnetar 1RXS J1708-4009. Results obtained without including vacuum birefringence in the magnetosphere are shown on the left column (labeled  \textit{VB off}). The middle and right columns respectively correspond to the intensity and polarization information obtained including vacuum birefringence in the magnetosphere with the phase shift $\Delta \phi$ set to zero (labeled \textit{VB on, $\Delta \phi = 0$}) and $\Delta \phi$ randomized in the range $[0, 2 \pi]$ (labeled \textit{VB on, $\Delta \phi =$ rnd}). The number of recorded photons is $\mathcal{N}_{\rm rec} = 10^8$. The inclination angle is $\alpha = 60^{\circ}$.}
	\label{fig:skymaps-J1708-alpha60-cap30}
\end{figure}

\subsection{Impacts of vacuum birefringence in the magnetosphere}
\label{sec:polarization}

\begin{figure}[t]
	\centering
	\includegraphics[width=1.0\linewidth]{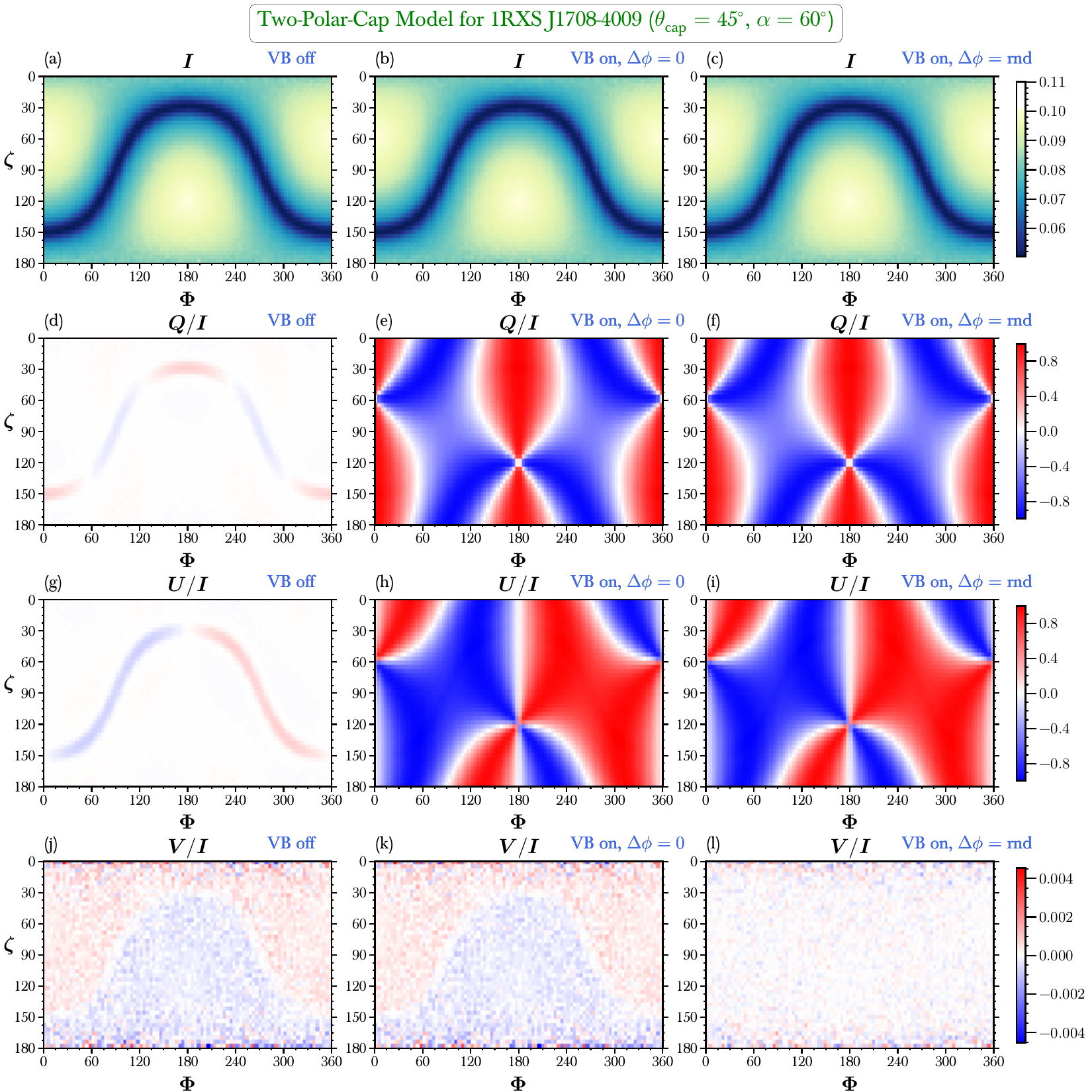}
	\caption{Same as Figure~\ref{fig:skymaps-J1708-alpha60-cap30}, but with $\theta_{\rm cap} = 45^{\circ}$.}
	\label{fig:skymaps-J1708-alpha60-cap45}
\end{figure}

The focus now turns to illustrating the effects of vacuum birefringence in the magnetosphere on the spectra and polarization signals emitted from NS surfaces. In Figure~\ref{fig:skymaps-J1708-alpha60-cap30}, we display the intensity $I$ and polarization ($Q/I$, $U/I$, $V/I$) sky maps obtained from models with two uniform antipodal polar caps extending from the respective magnetic poles to colatitudes $\theta_{\rm cap} = 30^{\circ}$, as functions of $\zeta$ (y-axis) and $\Phi$ (x-axis).   For comparison, Figure~\ref{fig:skymaps-J1708-alpha60-cap45} provides an identical display array but for the larger polar cap size of $\theta_{\rm cap} = 45^{\circ}$.  Note that a larger magnetic inclination angle of $\alpha = 60^{\circ}$ is chosen as an illustrative example that complements the intensity fitting solution in Section~\ref{sec:intensity}. Three scenarios are considered: (1) without VB in the magnetosphere (labeled ``VB off'', left column); (2) with VB in the magnetosphere, and the phase shift between the two polarization modes, $\Delta \phi$, is set to zero (labeled ``VB on, $\Delta \phi = 0$'', middle column); and (3) with VB acting in the magnetosphere, and a random phase approximation is adopted for $\Delta \phi$ (labeled ``VB on, $\Delta \phi = \rm rnd$'', right column), which is appropriate for the magnetospheric propagation. Comparing (2) and (3) will provide insights into the impact of the phase shift between the two polarization modes. The recoupling radius, $r_{\rm rec}$, is calculated using Equation~(\ref{eq:r_rc2}) with the polar field of \teq{B_p = 9.3 \times 10^{14}}Gauss as measured by an observer at infinity. 

As predicted by Equation~(\ref{eq:IQUV_rec}), the radiation intensity remains constant during the evolution of the polarization vector; see panels (a)-(c) in the top row.  This invariance is expected as the vacuum contribution to the dielectric tensor is non-dissipative, as noted by \cite{Adelsberg-2006-MNRAS}; absorptive contributions manifested through the creation of electron-positron pairs do not appear until above the \teq{2m_ec^2\sim 1}MeV threshold \citep{Erber-1966-RvMP,Tsai-1975-PhRvD}. 
In contrast, VB effects in the magnetosphere significantly enhance $Q/I$ and $U/I$, as shown in panels (d)-(i) of Figure~\ref{fig:skymaps-J1708-alpha60-cap30}. In the magnetar domain, $\omega /\wcyc \ll 1$, the emergent radiation exhibits strong linear polarization, with O-mode photons being dominant, across all magnetic colatitudes, as discussed in Section~\ref{subsec:magnetar}. In the absence of VB in the magnetosphere, the electric field vectors of photons emanating from different regions on the NS surface, corresponding to various magnetic field directions, tend to cancel each other, resulting in  depolarization at infinity. For $\theta_{\rm cap} = 30^{\circ}$, we can observe that $Q/I$ (panel (d)) and $U/I$ (panel (g)) vary between 0 and $\sim \pm 0.6$, so that the geometric cancellation is significant but not total.  For $\theta_{\rm cap} = 45^{\circ}$, the linear PD further decreases, $\Pi_l = \sqrt{(Q/I)^2 + (U/ I)^2} \lesssim 10{\%}$,  due to the superposition of a large
range of field directions spanned across a much larger cap; see panels (d) and (g) in Figure~\ref{fig:skymaps-J1708-alpha60-cap45}, comparing with the corresponding panels of Figure~\ref{fig:skymaps-J1708-alpha60-cap30}.  

It is notable that when VB is switched on, comparing these two figures clearly illustrates very little difference in the Stokes \teq{Q/I} and \teq{U/I} maps as $\theta_{\rm cap}$ changes, a hallmark of the action of VB in the magnetosphere.  In this situation, the electric field vectors of photons emerging from the surface rotate with the magnetic field direction until they reach the recoupling limit. For magnetars, since the recoupling happens significantly far from the NS surface, the GR effects are no longer important; see Figure~\ref{fig:fig-recoupling}. Thus, $\hat{\boldsymbol{k}}_{\rm GR}(r = r_{\rm rec}) = \hat{\boldsymbol{k}}_{\infty} $, i.e., the photon propagation vector remains unchanged as it travels from $r = r_{\rm rec}$ to $r \rightarrow \infty$.  As a result, the electric field vector is effectively frozen, $\boldsymbol{\mathcal{E}}_{\infty} = \boldsymbol{\mathcal{E}}_{\rm GR}(r = r_{\rm rec})$.  This means that for photons traveling in the same direction, defined by ($\zeta, \Phi$), their electric field vectors align, thereby preserving the high PD emergent at the surface \citep{Heyl2002}.  From Equation~(\ref{eq:IQUV_rec}), it follows that $Q/I \approx \cos2\Phi_{ \rm B}$ and $U/I \approx \sin2\Phi_{ \rm B}$ when $|\mathcal{E}_{\rm X, S}| \approx 0$, and therefore, these two Stokes parameters are independent of the phase shift $\Delta \phi$. Moreover, 
with $\Phi_{ \rm B}$ given in Equation~(\ref{eq:tanphiB_2}), $Q/I$ and $U/I$ observed at infinity depend solely on the geometric parameters, namely $\alpha$ and $\zeta$, as well as the rotational phase, $\Phi  = \Omega t$. This elucidates why the sky maps
of $Q/I$ and $U/I$ obtained including VB in the magnetosphere for the two $\theta_{\rm cap}$ values are apparently identical; see panels (e), (f), (h), (i) in Figures~\ref{fig:skymaps-J1708-alpha60-cap30}--\ref{fig:skymaps-J1708-alpha60-cap45}. 
The circular polarization $V/I$ is displayed in the last row of Figures~\ref{fig:skymaps-J1708-alpha60-cap30} and \ref{fig:skymaps-J1708-alpha60-cap45}. 
As previously discussed in \citet{Hu-2022-ApJ}, $V$ is small in the high-field domain. Furthermore, Equation~(\ref{eq:IQUV_rec}) indicates that if there is no phase lag between the two polarization modes, $\Delta \phi = 0$, then $V$ is not influenced by VB in the magnetosphere. Under a random phase approximation, $V = 0$, as is evident in Equation~(\ref{eq:IQUV_rec}).

\begin{figure}[t]
    \centering
    \includegraphics[width=1.0\linewidth]{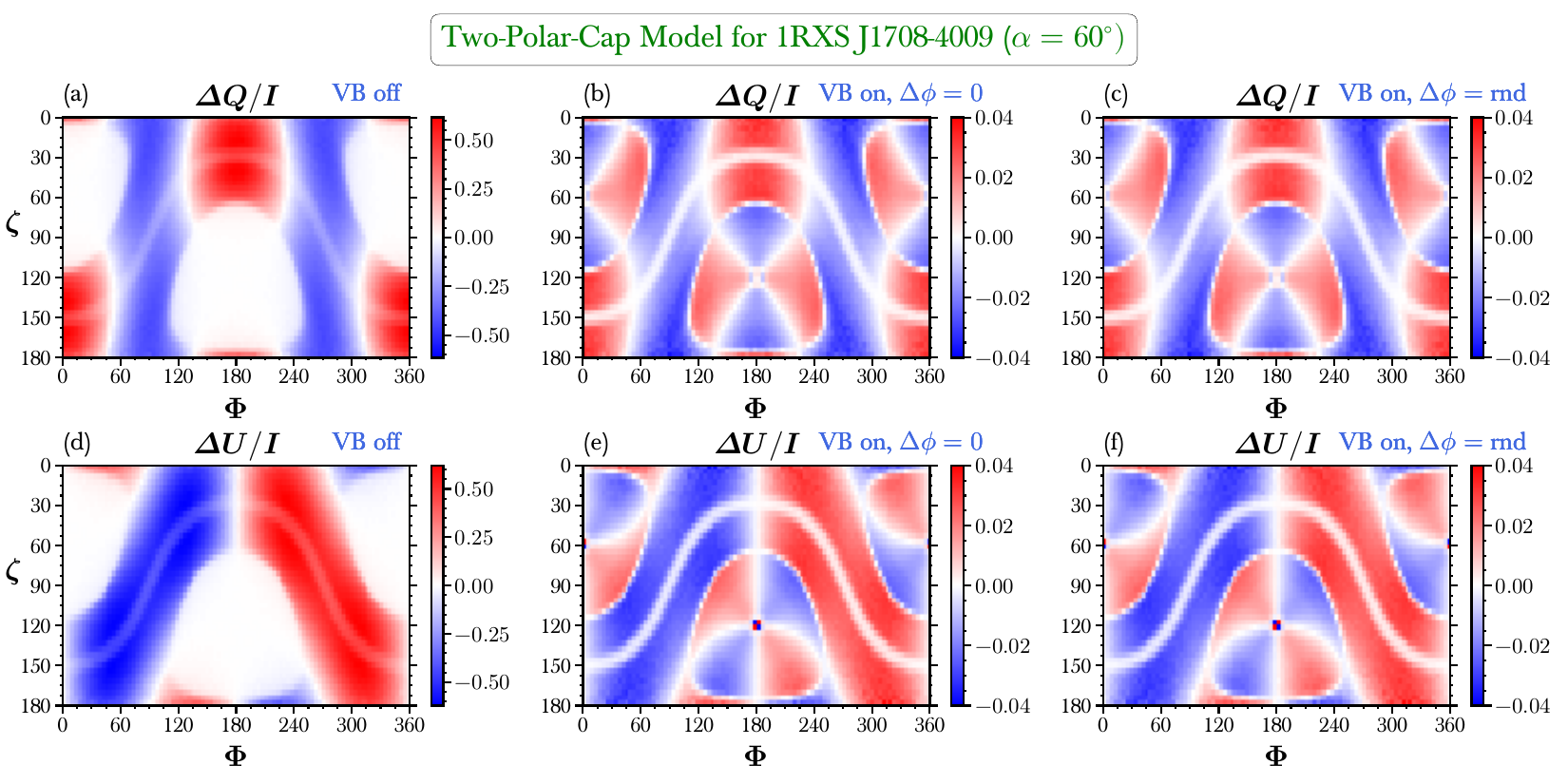}
    \caption{\edithere{Sky maps of the differences in $Q/I$ and $U/I$ between $\theta_{\rm cap} = 30^{\circ}$  (panels (d)- (i), Figure~\ref{fig:skymaps-J1708-alpha60-cap30}) and $\theta_{\rm cap} = 45^{\circ}$ (panels (d)- (i), Figure~\ref{fig:skymaps-J1708-alpha60-cap45}).  When vacuum birefringence is active in the magnetosphere (the center and right columns), the differences $\Delta Q/I$ and $\Delta U/I$ are at most a few percent.}}
    \label{fig:diff-skymaps-J1708-alpha60}
\end{figure}

\edithere{To highlight the similarities and differences in the polarization information in Figures~\ref{fig:skymaps-J1708-alpha60-cap30} and~\ref{fig:skymaps-J1708-alpha60-cap45}, we present in Figure~\ref{fig:diff-skymaps-J1708-alpha60} the sky maps of the differences in $Q/I$ (top row) and $U/I$ (bottom row) between $\theta_{\rm cap} = 30^{\circ}$ (Figure~\ref{fig:skymaps-J1708-alpha60-cap30}) and $\theta_{\rm cap} = 45^{\circ}$ (Figure~\ref{fig:skymaps-J1708-alpha60-cap45}). It is evident that when VB is on (middle and right columns), the difference between the two capsizes is small, typically less than around $4\%$, highlighting the impact of VB in action.  In contrast, when VB is switched off, there are large differences between these polarization measures for the two cap sizes (left column).}

\begin{figure}[t]
	\centering
	\includegraphics[width =1.0\linewidth]{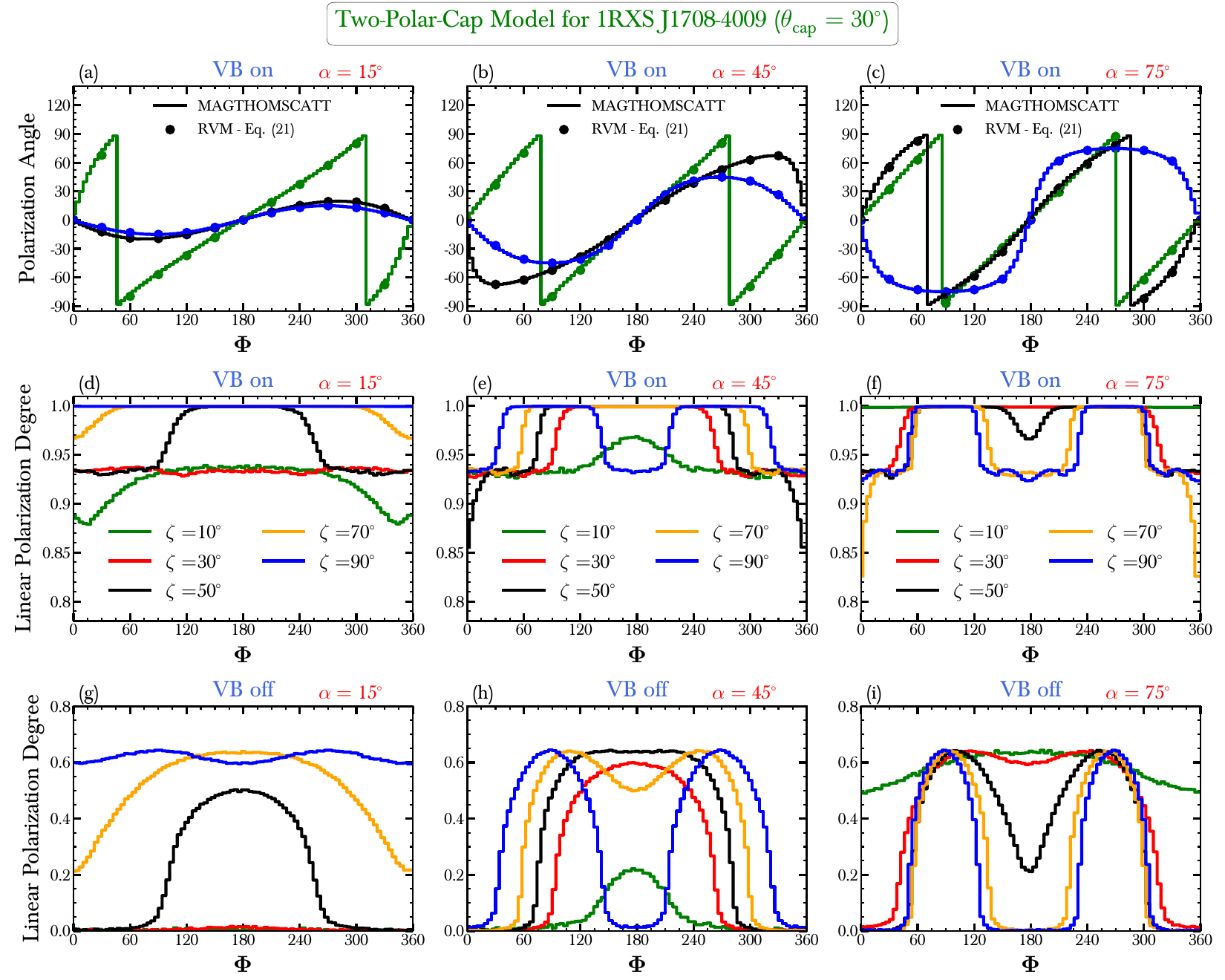}
	\caption{Top row: simulated pulse profiles (solid curves) for the PA as a function of rotational phase \teq{\Phi = \Omega t} (in degrees) for three viewing angles: $\zeta = 10^{\circ}$ (green), $\zeta = 50^{\circ}$ (black), $\zeta = 90^{\circ}$ (blue).  The corresponding RVM PA traces, obtained from Equation~(\ref{eq:tanphiB_2}), are shown with points that are color-coded as for the simulated ones.  Middle and bottom rows: simulated linear PD pulse profiles with the inclusion of VB in the magnetosphere (middle row) and without VB (bottom row), for five selected viewing angles $\zeta = 10^{\circ},30^{\circ}, 50^{\circ}, 70^{\circ}, 90^{\circ}$. The simulated results are obtained from models for the magnetar 1RXS J1708-4009 that employ two uniform antipodal polar caps extending from the respective magnetic poles to colatitudes $\theta_{\rm cap} = 30^{\circ}$. } 
	\label{fig:fig-PD-J1708}
\end{figure}

The $Q$ and $U$ results shown in Figures~\ref{fig:skymaps-J1708-alpha60-cap30}--\ref{fig:skymaps-J1708-alpha60-cap45} indicate that in the presence of VB in the magnetosphere, the PA of X-ray emission from magnetars can be approximated by a rotating vector model (RVM), $\chi \equiv 1/2\arctan{(U/Q)} \approx \Phi_B \equiv \chi_{\rm RVM}$.  We note that the expression of the RVM PA given by Equation~(\ref{eq:tanphiB_2}) corresponds to the convention where PA increases counterclockwise in the sky, a convention commonly used in observations, see e.g.,~\cite{Tong-2021-mnras, Everett-2001-ApJ} and references therein.  \edithere{To illustrate a variety of light curve and phase-resolved polarization possibilities}, the top row of Figure~\ref{fig:fig-PD-J1708} presents the simulated PA pulse profiles (solid curves) for 1RXS J1708-4009 using $\theta_{\rm cap} = 30^{\circ}$  as a function of rotational phase for three selected viewing directions: $\zeta = 10^{\circ}$ (green), $\zeta = 50^{\circ}$ (black), and $\zeta = 90^{\circ}$ (blue).  The three columns correspond to three values of the magnetic inclination angle: $\alpha =$ $15^{\circ}$ (left column), $45^{\circ}$ (middle column), and $75^{\circ}$ (right column). The 180-degree swing from $\chi = 90^{\circ}$ to $\chi = -90^{\circ}$ observed in some cases corresponds to the transition from $U > 0$ to $U < 0$ when $Q<0$. For $Q>0$, the sign change in $U$ from negative to positive leads to a transition in PA from $\chi <0$ to $\chi >0$ at $\Phi = 180^{\circ}$. At large viewing angles ($\zeta \geq 50^{\circ}$) and small inclination angles ($\alpha \leq 45^{\circ}$), the PA varies only moderately with the rotational phase. \edithere{This latter behavior is consistent with the PA pulse profile observed for this source \citep{Zane-2023-ApJ}, and also for that in soft X rays below around 4 keV from 1E 1841-045 \citep{Stewart-2025-ApJL,Rigoselli-2025-ApJ}.}  As previously mentioned, if VB is present, the RVM PA in Equation~(\ref{eq:tanphiB_2}) should represent a good approximation to the simulated one in the magnetar domain.  To verify this, we show in the top row of Figure~\ref{fig:fig-PD-J1708} the RVM PA results obtained from Equation~(\ref{eq:tanphiB_2}), displayed using points that are color-coded as for the simulated ones.  The simulated and RVM PA values are essentially indistinguishable, confirming the validity of our VB implementation.
 
The middle and bottom rows of Figure~\ref{fig:fig-PD-J1708} displays the linear polarization pulse profiles for 1RXS J1708-4009 using $\theta_{\rm cap} = 30^{\circ}$ for five selected viewing angles, $\zeta = 10^{\circ}, 30^{\circ}, 50^{\circ}, 70^{\circ}, 90^{\circ}$.  Without VB effects in the magnetosphere (bottom panels), the linear PD varies between 0 and 60$\%$, the same range as is realized in {\sl IXPE} observations \citep{Zane-2023-ApJ}.  \edithere{This PD range is also approximately commensurate with that seen by {\sl IXPE} in 1E 1841-045 \citep{Stewart-2025-ApJL,Rigoselli-2025-ApJ}.  In strong contrast}, when the influence of VB is included, the linear PD significantly increases (middle row), with values exceeding $80\%$. This is notably higher than the observed PD for 1RXS J1708-4009, where $\Pi_l < 50\%$ for $E_{\gamma} < 4$~keV; see Figure~4 in \citet{Zane-2023-ApJ}.  This result suggests that when VB is switched on for magnetospheric propagation, the radiation emerging at the surface must possess lower linear PDs than those that are predicted by our current atmospheric model.

To explain the modest polarization of 1RXS J1708-4009, \cite{Zane-2023-ApJ} proposed a model comprising a hot atmospheric patch and a warm condensed region.  \edithere{A similar condensed surface premise was invoked by \cite{Rigoselli-2025-ApJ} to explain the moderate PD below around 4 keV detected by {\sl IXPE} in 1E 1841-045.  Recently,} with a particle bombardment model, \citet{Kelly2024a} found a phase-averaged PD compatible with that observed for 1RXS J1708-4009 by {\sl IXPE} in the $2-3$ keV energy range. Nevertheless, they concluded that to fully account for the polarization features observed by {\sl IXPE} in magnetar sources, the particle bombardment model would need to be combined with other emission scenarios, such as atmospheric models, resonant Compton scattering, and condensed surface models. Alternatively,  spectral proximity to the vacuum resonance in NS atmospheres \citep{Lai-Ho-2003-ApJ, Lai-2023-PNAS} could lead to depolarization.  The vacuum resonance is a narrow region of anomalous dispersion that arises due to the competition between vacuum and plasma birefringence deep within the atmosphere where the densities are higher.  At the resonance, a photon can convert from one polarization mode to the other; therefore, the overall surface signals are expected to depolarize. \cite{Ho2003} considered the two limiting cases at the vacuum resonance, namely no mode conversion and complete mode conversion, whereas partial mode conversion has been explored by \cite{Adelsberg-2006-MNRAS, VanAdelsberg2009, Lai-2023-PNAS, Kelly2024b}; see also references therein. Assuming a constant mode conversion probability of $\sim 30\%$ near the surface, we can reproduce the phase-averaged PD measured for 1RXS J1708-4009. Nonetheless, in reality, the conversion probability depends on the photon direction with respect to the magnetic field and its energy \citep{Lai-Ho-2003-ApJ}. Therefore, to accurately capture the phase-resolved polarization properties, a rigorous implementation of mode conversion is needed; this will be the focus of a future extension of {\sl MAGTHOMSCATT}.  This enhancement will allow us to disentangle the energy-dependent, atmospheric vacuum birefringence contributions to observed polarizations from the achromatic magnetospheric VB influences. 

\begin{figure}[t]
	\centering
	\includegraphics[width=1.0\linewidth]{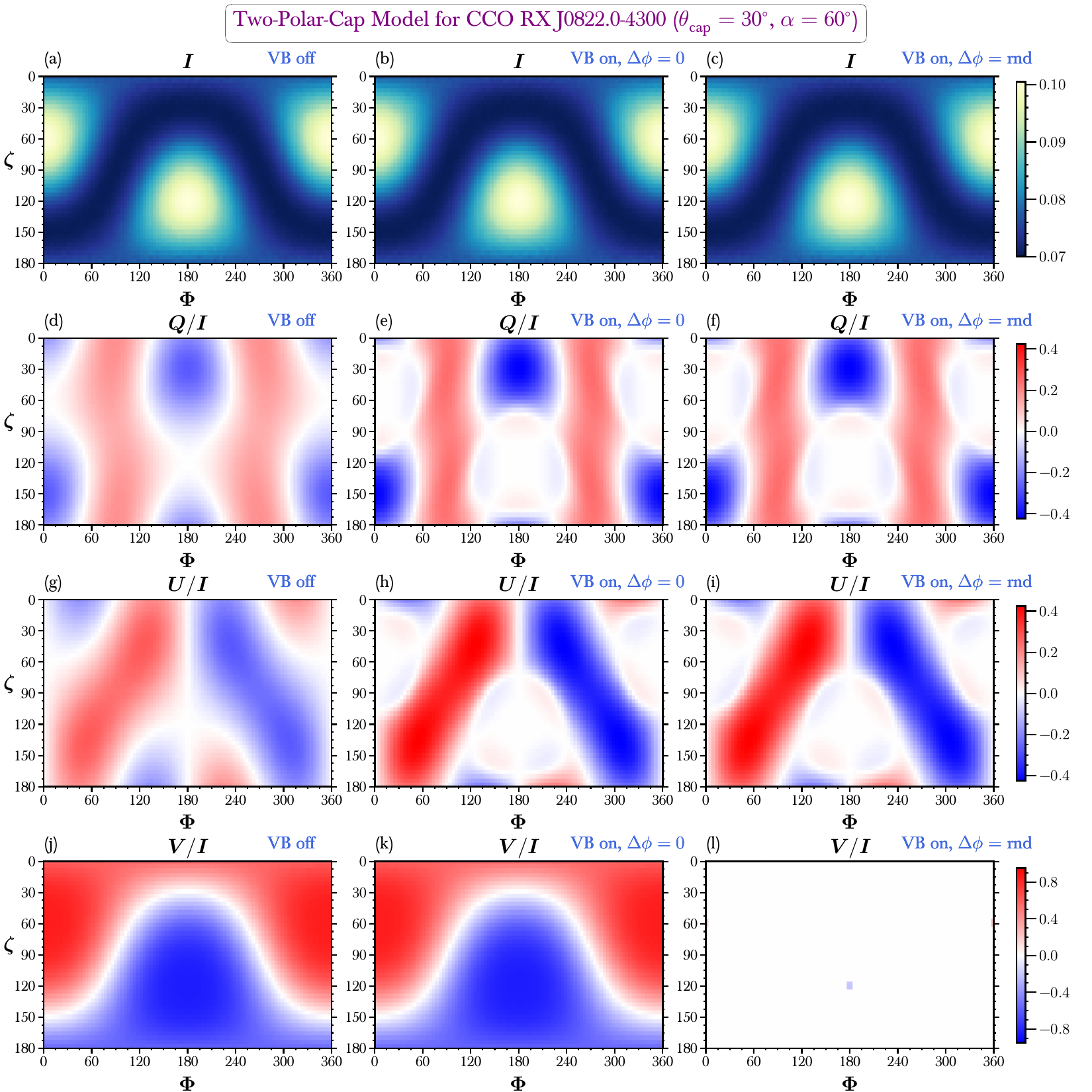}
	\caption{Intensity and polarization ($Q/I$, $U/I$, $V/I$) sky maps obtained from the  modelings of two uniform antipodal polar caps extending from the respective magnetic poles to colatitudes $\theta_{\rm cap} = 30^{\circ}$ for CCO RX~J0822.0-4300 with temperature $kT = 0.42$~keV and $E_{\gamma} = 1-5$ keV as observed at infinity. Results obtained without including vacuum birefringence in the magnetosphere are shown on the left column (labeled  \textit{VB off}). The middle and right columns respectively correspond to the intensity and polarization information obtained including vacuum birefringence in the magnetosphere with the phase shift $\Delta \phi$ set to zero (labeled \textit{VB on, $\Delta \phi = 0$}) and $\Delta \phi$ randomized in the range $[0, 2 \pi]$ (labeled \textit{VB on, $\Delta \phi =$ rnd}). The number of recorded photons is $\mathcal{N}_{\rm rec} =  5\times 10^9$. The inclination angle is $\alpha = 60^{\circ}$.}
	\label{fig:fig-skymaps-CCO-alpha60-cap30}
\end{figure}

\begin{figure}[t]
	\centering
	\includegraphics[width=1.0 \linewidth]{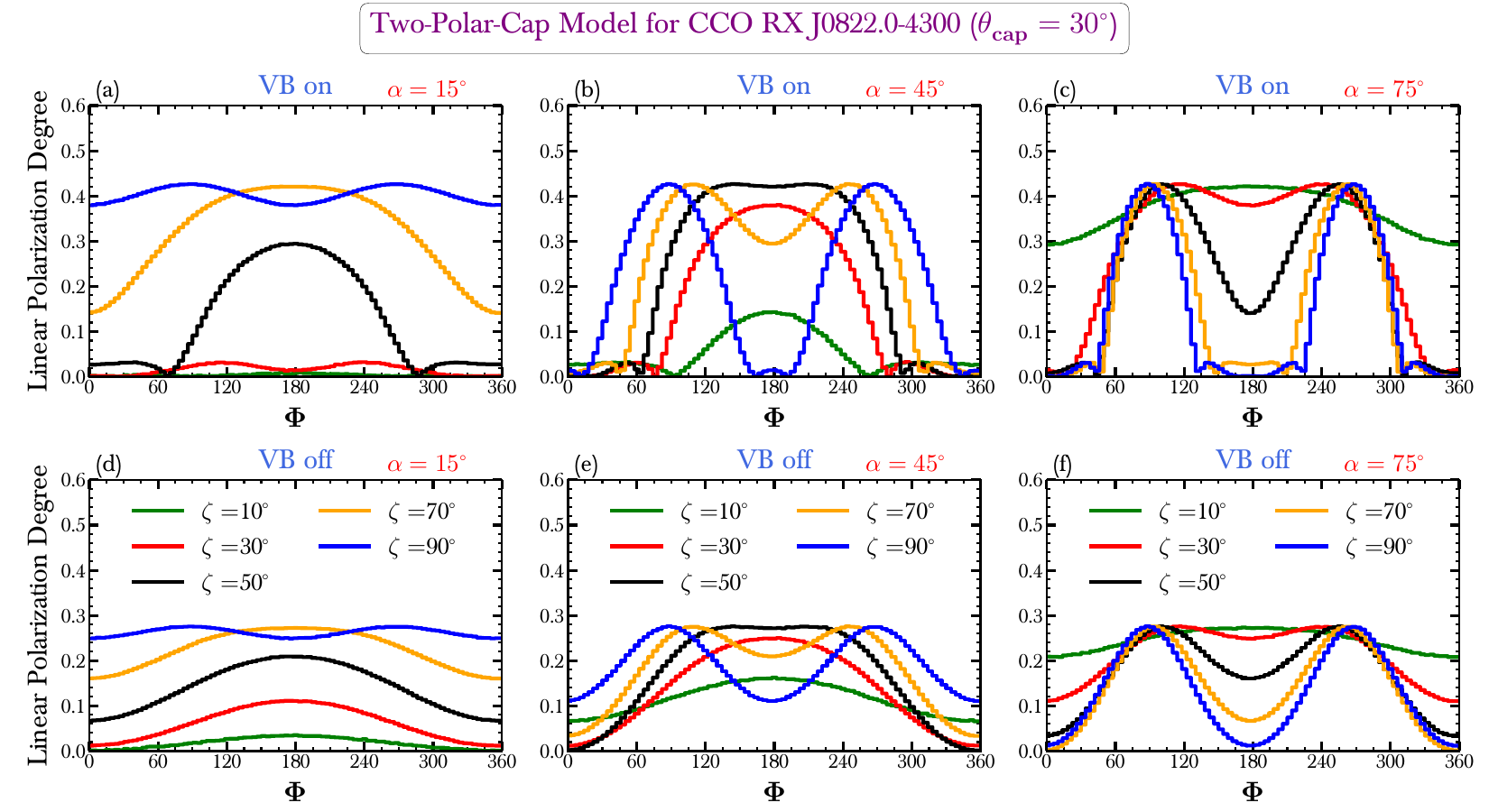}
	\caption{Simulated pulse profiles for the linear PD  obtained for two uniform antipodal polar caps extending from the respective magnetic poles to colatitudes $\theta_{\rm cap} = 30^{\circ}$ for   CCO RX~J0822.0-4300 with temperature $kT = 0.42$~keV and $E_{\gamma} = 1-5$ keV as observed at infinity. These pulse profiles plotted as functions of the rotation phase $\Phi$ for five selected viewing angles $\zeta = 10^{\circ},30^{\circ}, 50^{\circ}, 70^{\circ}, 90^{\circ}$. The results shown on the top (bottom) are obtained with (without) including VB in the magnetosphere.}
	\label{fig:fig-PD-CCO}
\end{figure}

The impact of VB in the magnetosphere is less pronounced for NSs with lower magnetizations. To illustrate this, in Figures~\ref{fig:fig-skymaps-CCO-alpha60-cap30} and \ref{fig:fig-PD-CCO}, we respectively show model the ($I$, $Q/I$, $U/I$, $V/I$) sky maps and PD pulse profiles suited for the CCO pulsar RX~J0822.0-4300 with temperature $kT = 0.42$~keV and $E_{\gamma} = 1-5$ keV as observed at infinity \citep{Gotthelf2010}.  The polar field strength, as measured by an observer at infinity, is \teq{5.7\times 10^{10}}Gauss, corresponding to a polar field strength of \teq{8.5 \times 10^{10}}Gauss in the LIF at the surface.  Therefore, the soft X-ray data are somewhat above the cyclotron energies, corresponding to frequency ratios at the magnetic pole in the LIF sampled for RX~J0822.0-4300 realizing values $1.3 \lesssim \omega / \wcyc \lesssim 6.7$.  In this case, the emergent linear PDs are moderate and $X$-mode photons dominate. Given this low magnetic field strength, \teq{B_p = 5.7 \times 10^{10}}Gauss, the two modes recouple at a few stellar radii from the surface, $r_{\rm rec} \sim 4 R_{\rm NS}$. Figure~ \ref{fig:fig-skymaps-CCO-alpha60-cap30} shows that the radiation intensity is not influenced in the magnetosphere, as evident from Equation~(\ref{eq:IQUV_rec}). Under the random phase approximation, the magnetospheric QED effects wash out the strong circular polarization at the surface. This is a clear prediction for such modestly magnetized stars that could motivate the development of circular polarization detection technologies in X-rays, a difficult enterprise.

For the linear polarization, magnetospheric QED effects enhance both $Q$ and $U$, albeit only moderately: see Figure~\ref{fig:fig-skymaps-CCO-alpha60-cap30}; observe also the morphological similarity between the left column of panels (VB off) and the right two columns (VB on).  The maximum linear PD then increases from $\sim 30\%$ to $\sim 40\%$, which is non-negligible; see Figure~\ref{fig:fig-PD-CCO}.  Yet this enhancement is much more modest than that for the 1RXS J1708-4009 magnetar case, suggesting that CCOs are not optimal targets for searching for evidence of vacuum birefringence in their magnetospheres.  Adding to this, RX~J0822.0-4300 is fainter than typical magnetars, rendering X-ray polarimetry with {\sl IXPE} a more challenging prospect.  Here, we do not illustrate light curve fitting for RX~J0822.0-4300 as it inherently involves modeling hot spots of different temperatures \citep{Alford-2022-ApJ}; such a task is detailed in the upcoming lengthy study of Dinh Thi et al. (in prep.).

\newpage

\vphantom{p}

\newpage

\section{Conclusions}
 \label{sec:conclusions}
Analyses of X-ray emission from NS surfaces can provide valuable insights into their interior and surface properties as well as constraints on their geometric configurations. In this paper, we study X-ray emission from extended surface regions using our Monte Carlo simulation of atmospheric radiative transport, {\sl MAGTHOMSCATT}, which keeps track of the trajectory and complex electric field vector for each photon. The general relativistic propagation of light from the stellar surface to infinity is included, and our presentation details how we model the effects of vacuum birefringence in the magnetosphere on the polarization information.  As an illustration of the simulation's utility, we modeled the intensity pulse profile of magnetar 1RXS J1708-4009, observed in the energy range of 0.5 to 3 keV with XMM,  assuming emission configurations of two identical antipodal isothermal polar caps. 
\edithere{Nine different sizes for polar caps are considered, $\theta_{\rm cap}$ ranges from $5^{\circ}$ to $45^{\circ}$ in increments of $5^{\circ}$. Each light curve model is defined by $(\theta_{\rm cap}, \alpha, \zeta)$.  We found the best fit to the full intensity light curve data corresponds to $(\theta_{\rm cap}, \alpha, \zeta) = (40^{\circ}, 80^{\circ}, 10^{\circ})$.}
Nevertheless, there are multiple parameter sets that also yield comparably good fits to the intensity data. While these models can satisfactorily reproduce the overall shape of the observed intensity pulse profile,  due to the azimuthal symmetry of the polar caps, the small asymmetric structure observed in the light curves at phase $\sim 0.5$ cannot be captured. \edithere{Excising the data points between phases 0.20 and 0.65 significantly improves the fit, and the parameter distributions shift towards lower $\alpha$ and $\zeta$. The best fit to the excised intensity data corresponds to $(\theta_{\rm cap}, \alpha, \zeta) = (15^{\circ}, 50^{\circ}, 10^{\circ})$. This underscores the need for a more detailed analysis of the hot-spot shapes and localations. } Our {\sl MAGTHOMSCATT} simulation can be adapted to treat more complex surface emission configurations, such as inhomogeneous temperature distribution and asymmetric shapes of hot spots. Such investigations are deferred to future work, with the possible incorporation of machine learning techniques.

The parameter degeneracy present when using only intensity data can potentially be resolved by incorporating additional information from polarization measurements, which are now available for some magnetars with {\sl IXPE} observations.  As a photon leaves the NS surface and propagates through the birefringent vacuum induced by the magnetospheric \teq{\boldsymbol{B}} field, its electric field vector rotates adiabatically with the local direction of \teq{\boldsymbol{B}}. We have incorporated in our {\sl MAGTHOMSCATT} simulation the evolution of the electric field vector due to this effect.  The recoupling surface, which depends on photon energy, propagation direction, and magnetic field strength, was calculated for two representative cases: (1) magnetar 1RXS J1708–4009 and (2) CCO RX J0822.0–4300. For the former, $r_{\rm rec}$ varies between 0 at the magnetic pole to $\sim 200 R_{\rm NS}$ at the magnetic equator, while for the CCO, the recoupling happens much closer to the stellar surface.  In the absence of a treatment of the impact of the vacuum resonance deep in the atmosphere, which arises when the contributions of the magnetized vacuum and plasma to the dielectric tensor tend to cancel each other, radiation emanating from magnetar surfaces is strongly polarized. Such high polarization is preserved through adiabatic evolution of the electric field vector in the magnetosphere, leading to significant PDs at infinity.  For NS of lower magnetizations, such as CCO  RX J0822.0–4300, the effect on linear PD is much less pronounced, where the maximum linear PD increases by $\sim 10\%$. In addition, the birefringence influences on the circular polarization in this more weakly magnetized example are strongly dependent on the phase lag between the two polarization modes; a more rigorous analysis is needed to explore such physics manifestations more incisively.  In future extensions of {\sl MAGTHOMSCATT}, we will focus on incorporating the vacuum resonance physics in the atmosphere and quantifying its effects on the polarization properties of surface signals at different X-ray energies.

\begin{acknowledgments}

\vspace{-5pt}
\centerline{Acknowledgments}
\vspace{8pt}

\edithere{The authors thank the anonymous referee and Wynn Ho for suggestions helpful to the polishing of the manuscript.}
M.G.B. thanks NASA for generous support under awards 80NSSC24K0589, 80NSSC25K7257 and 80NSSC25K0079.  
This work was supported in part by the Big-Data Private-Cloud Research Cyberinfrastructure MRI-award funded by NSF under grant CNS-1338099 and by Rice University's Center for Research Computing (CRC). The material is based upon work for which G.Y. is supported by NASA under award number 80GSFC24M0006.

\end{acknowledgments}

%

\vspace{5mm}





\vspace{-15pt}
\appendix
\vspace{-15pt}

\section{General relativistic light propagation and dipolar field morphology}
 \label{sec:GRprop}
We adopted the approximation put forward by \citet{Poutanen2020}  to describe the light bending effect in the framework of the Schwarzschild metric for slowly rotating NSs. \edithere{This approximation is accurate to the level of $0.01\%$ for the NS mass and radius considered in this work.} The trajectory plane of a photon is defined by its wavevector $\boldsymbol{k}_{\rm S}$ at the emission point and the local radial vector $\boldsymbol{r}_{\rm S}$; see Figure~3 in \cite{Hu-2022-ApJ}.  The angle $\theta_{zr} = \theta_{zr}(\Psi)$ between the photon momentum vector $\boldsymbol{k}_{\rm GR}$ and the local radial vector $\boldsymbol{r}$ at any altitude is given by
\begin{equation}
	\cos \theta_{zr} \; = \; \frac{1}{\Psi_b} \sqrt{\Psi_b^2 - \Psi^2(1 - \Psi)} \quad  , \quad
	\Psi_b  \; = \; \frac{\Psis \sqrt{1 - \Psis}}{\sin{\theta_z}} \quad , 
\end{equation}
as in Equation~(B6) in \citet{Hu-2022-ApJ}. Herein, $\Psi = \rSch/r$, $\Psis= \rSch/R_{\rm NS}$,\teq{\Psi_b = \rSch/b} for an impact parameter \teq{b}, defined as the perpendicular distance between the photon's trajectory at infinity and the parallel line that passes through the NS center.  At the surface, $\theta_{zr}(\Psis ) = \theta_z $, as expected. The angle $\eta$ between $\boldsymbol{r}$ and the direction to the observer $\hat{\boldsymbol{k}}_{\infty} = \hat{\boldsymbol{r}}_{\infty}$ relates to $\theta_{zr}$ through Equation~(2) in \citet{Poutanen2020}.
The unit position and propagation vectors, $\hat{\boldsymbol{r}}$ and $\hat{\boldsymbol{k}}_{\rm GR}$, can be expressed in terms of $\hat{\boldsymbol{k}}_{\infty}$ and $\hat{\boldsymbol{j}}_{\infty}$ given Equation~(9) in \citet{Hu-2022-ApJ}, as
\begin{equation}
	\hat{\boldsymbol{r}} \; =  \; \cos\eta\, \hat{\boldsymbol{k}}_{\infty} +  \sin \eta\,  \hat{\boldsymbol{j}}_{\infty}  \quad , \quad
	\hat{\boldsymbol{k}}_{\rm GR} \; = \; \cos(\eta - \theta_{zr})\,  \hat{\boldsymbol{k}}_{\infty} +  \sin (\eta-\theta_{zr})\,  \hat{\boldsymbol{j}}_{\infty} \quad . \label{eq:ray-tracing}
\end{equation}
The details of this trajectory information are expounded at length in \cite{Hu-2022-ApJ}, as is the treatment of parallel transport of photon electric field vectors.

The magnetic dipole field in the curved spacetime of the Schwarzschild metric that is used in {\sl MAGTHOMSCATT} was formulated by \citet{Wasserman1983}, see also \citet{Gonthier1994, Story2014}. In the polar coordinates defined by the magnetic dipole moment axis $\hat{\boldsymbol{\mu}}_{\rm B}$, the magnetic field in the LIF is expressed as
\begin{equation}
  \boldsymbol{B}  \; = \; 3 \frac{B_p R_{\rm NS}^3}{\rSch^3}\Psi^3
   \Bigl\{  \xi_r(\Psi)\cos\theta \; \hat{\boldsymbol{r}} + \xi_\theta (\Psi)\sin\theta \; \hat{\boldsymbol{\theta}}
   \Bigl\}  \quad , 
 \label{eq:BGR}
\end{equation}
in which $B_p$ is polar magnetic field strength as inferred at infinity, while $\xi_r(\Psi)$ and $\xi_{\theta}(\Psi)$ are given in Equation~(B2) in \citet{Hu-2022-ApJ}. As $\Psi \rightarrow 0$, 
\teq{\xi_r(\Psi )\approx 1/3} and \teq{\xi_{\theta} (\Psi )\approx 1/6} and Equation~(\ref{eq:BGR}) reduces to the a flat spacetime dipole field, as expected.
\section{Derivation of the recoupling radius}
\label{sec:appendix}

This Appendix details the derivations of the expressions for the recoupling radius that were obtained in \citet{Heyl-2000-MNRAS} and \citet{Adelsberg-2006-MNRAS} and that are highlighted in Section~\ref{sec:recoupling}.

\subsection{Heyl and Shaviv's approach} \label{subsec:r_rec_HS}
In a non-rotating stellar configuration, at $r \gg R_{\rm NS}$, the photon travels radially, and $\theta_{\rm kB}$, that is, the angle between the photon propagation vector \teq{\bf{\hat{k}}_{\infty}} and the local magnetic field direction\teq{\hat{\boldsymbol{B}}}, is then independent of altitude. 
The recoupling condition is defined in Equation~(\ref{eq:recoupling_1}). 
%
%
In this \teq{r\gg \rns} domain, the field geometry reduces to the familiar flat spacetime dipolar form:
\begin{equation}
	\boldsymbol{B} \; \approx \; \frac{1}{2}\frac{B_p R_{\rm NS}^3}{r^3}
	\Bigl\{ 2\cos\theta \, \hat{\boldsymbol{r}}_{\infty} + \sin\theta \, \hat{\boldsymbol{\theta}}_{\infty}
	\Bigl\} 
    \quad \Rightarrow\quad
    B \;\equiv\; \bigl|\boldsymbol{B}\bigr|  \; \approx \; \frac{1}{2}\frac{B_p R_{\rm NS}^3}{r^3}\sqrt{1 + 3\cos^2 \theta} \quad .
 \label{eq-apd:Binf}
\end{equation}
Taking the derivative of $B$ with respect to $r$ at $r = r_{\rm rec}$, one can obtain:
\begin{equation}
	\frac{1}{B (r = r_{\rm rec})} \left |\left. \frac{\partial B}{\partial r} \right|_{r = r_{\rm rec}}  \right| \; = \; \frac{3}{r_{\rm rec}} \label{eq-apd:dB} \quad .
\end{equation}
This constitutes the right hand side of Equation~(\ref{eq:recoupling_2}).  From Equation~(\ref{eq-apd:Binf}), the angle between the direction of propagation and the external field is quickly shown to satisfy
\begin{equation}
	\cos \theta_{\rm kB}  \; = \; {\bf{\hat{k}}}_{\infty} \cdot  \hat{\boldsymbol{B}} \; \approx \; \frac{ {\bf{\hat{r}}}_{\infty} \cdot {\boldsymbol{B}}}{B}
    \; = \; \frac{2\cos\theta}{\sqrt{1 + 3\cos^2\theta}} \quad \hbox{or}\quad
    \sin \theta_{\rm kB} \; \approx \; \frac{\sin\theta}{\sqrt{1 + 3\cos^2\theta}} \quad .
 \label{eq:theta_kB}
\end{equation}
Inserting these last two results into Equation~(\ref{eq:recoupling_2}) for the magnitude \teq{|\boldsymbol{\Omega}_{\rm B}(r = r_{\rm rec})| \equiv k\Delta n(r = r_{\rm rec})} of the birefringence vector, and employing Equation~(\ref{eq:deltan}) for \teq{\Delta n}, the governing equation for \teq{r_{\rm rec}} becomes
\begin{equation}
	\frac{\omega}{c}\frac{\alpha_{\rm f}}{30\pi} \left(\frac{B(r = r_{\rm rec}) }{B_{\rm cr}}\right)^2\sin^2 \theta_{\rm kB} \; =\; 
	\frac{\nu}{c}\frac{\alpha_{\rm f}}{15} \, \frac{B^2_p R_{\rm NS}^6}{4B_{\rm cr}^2 r_{\rm rec}^6} \, \sin^2 \theta    \; = \;  \frac{3}{r_{\rm rec}} \quad ,
 \label{eq:birefringence_vec_r_rec}
\end{equation}
%
%
where $\nu = \omega / (2\pi)$ is the photon frequency.  Rearranging this result, the recoupling radius can be expressed as 
\begin{equation}
	r_{\rm rec} \; =  \; \left( \frac{\nu}{c}\frac{\alpha_{\rm f}}{45}\right)^{1/5}\left(\frac{\mu_{\rm B} \sin\theta}{B_{\rm cr}}\right)^{2/5} \quad ,
 \label{eq-apd:r_rc2}
\end{equation}
where $\mu_{\rm B} = B_pR^3_{\rm NS} / {2}$.
This expression for $r_{\rm rec}$ is identical to that in Equation~(4) of \citet{Heyl2002}. Using  $E_1 = {E_{\gamma}} / {1\mathrm{keV}}$ and $B_{12} = {B_p}/{(10^{12} G)}$, Equation~(\ref{eq-apd:r_rc2}) becomes:
\begin{equation}
	\frac{r_{\rm rec}}{R_{\rm NS}} \; \approx 11 \, E_1^{1/5} 
    (B_{12}\sin\theta)^{2/5} \quad , 
\end{equation}
which is equivalent to Equation~(\ref{eq:r_rc2}).

\subsection{Van Adelsberg and Lai's approach}

In \cite{Adelsberg-2006-MNRAS}, stellar rotation is of central importance to the specification of the recoupling radius.  In their work, the adiabatic condition and the recoupling (polarization limiting) radius are respectively via
\begin{equation}
     \bigl| {\bf{{\Omega}_B}} \bigr| \; \gg \;  2 \frac{d\Phi_{\rm B}}{ds}
     \quad \hbox{and}\quad
     \bigl| {\bf{{\Omega}_{B{,\rm pl}}}} \bigr| 
     \; = \;  2 \frac{d\Phi_{\rm B}}{dr} \quad .
 \label{eq:app-adiabatic_rpl_AL}
\end{equation}
Herein, $\Phi_{\rm B}$ is the angle between the component of the external field perpendicular to the line of sight and the x-axis.  Using the Cartesian coordinates $\hat{x}$ and $\hat{y}$ given in Equation~(\ref{eq:coordinate}), one can write $\tan\Phi_{\rm B}$ as:
\begin{equation}
    \tan \Phi_{\rm B} \; = \; \frac{\boldsymbol{B}_{r \gg R_{\rm NS}} \cdot \hat{y}}{\boldsymbol{B}_{r \gg R_{\rm NS}}\cdot \hat{x}} \quad , \label{eq:tanphiB}
\end{equation}
Let us choose a coordinate system where the unit NS rotation vector along the \teq{z}-axis: \teq{\hat{\boldsymbol{\Omega}} =  (0, 0, 1)}.  The line of sight $ \hat{\boldsymbol{k}}_{\infty}$ can be written as:
\begin{equation}
    \hat{\boldsymbol{k}}_{\infty} \; = \;  (\sin \zeta, 0, \cos \zeta) \quad , \label{eq:kinf_AL}
\end{equation}
where $\zeta$ is the viewing angle with respect to the rotational axis.  For a rotating NS with its unit magnetic dipole moment vector $\hat{\boldsymbol{\mu}}_{B}$ inclined at an angle $\alpha$ to the spin axis (i.e., \teq{\hat{\boldsymbol{\Omega}}}), simple rotations yield
\begin{equation}
    \hat{\boldsymbol{\mu}}_{B} \; = \; 
    \bigl( \sin\alpha \cos \Phi, \, \sin \alpha \sin \Phi, \, \cos\alpha \bigr) \quad ,
 \label{eq:muB_AL}
\end{equation}
at any time \teq{t} that specifies a rotational phase $\Phi = \Omega t$.  From Equations~(\ref{eq:kinf_AL}) and Equation~(\ref{eq:muB_AL}), we can write $\cos\theta$ in terms of $\alpha$, $\zeta$, and $\Phi$ at any instant as:
\begin{equation}
    \cos \theta =  \hat{\boldsymbol{k}}_{\infty} \cdot \hat{\boldsymbol{\mu}}_{B} =  \cos\alpha \cos\zeta + \sin\zeta \sin\alpha \cos \Phi.
 \label{eq:theta_AL}
\end{equation}
This agrees with Equation~(18) in \citet{Heyl-2000-MNRAS}. The second term on the right-hand side of Equation (\ref{eq:theta_AL})  exhibits an opposite sign compared to that in Equation (49) in \citet{Adelsberg-2006-MNRAS}. This again suggests that their definition of the inclination angle could  carry opposite sign to ours.  Using the various geometrical definitions, Equation~(\ref{eq:tanphiB}) becomes:
\begin{equation}
    \tan \Phi_{\rm B} \; = \; \frac{\sin \alpha \sin \Phi}{- \cos \alpha \sin \zeta + \cos \Phi \cos \zeta \sin \alpha} 
 \label{eq:app-tanphiB_2}
\end{equation}
This equation for the orientation angle of the field at high altitudes is consistent with Equation~(22) in \cite{Heyl-2000-MNRAS}. However, the sign of the first term in the denominator on the right-hand side of Eq.~(\ref{eq:app-tanphiB_2}) differs from that in Equation~(57) of \citet{Adelsberg-2006-MNRAS}, which could be due to some difference in the sign convention for the angles.  In addition, we note that if the emergent radiation is fully dominated by the O-polarization mode ($\mathcal{E}_{\rm X,S} = 0$), then Equation~(\ref{eq:IQUV_rec}) gives $Q/I = \cos 2\Phi_{ \rm B}$ and $U/I = \sin 2\Phi_{ \rm B}$. Therefore, the PA, $\chi = (1/2)\, \arctan{(U/Q)}$, coincides with $\Phi_{ \rm B}$, which is the expression of the PA in the RVM for dipole magnetic field geometry, see e.g., \citet{Tong-2021-mnras} and references therein. The RVM PA expression provided in Equation~(\ref{eq:app-tanphiB_2}) follows the convention that PA increases counterclockwise on the sky. This differs by a negative sign from the RVM PA that increases in the clockwise direction, as previously noted by \cite{Everett-2001-ApJ}. The counterclockwise convention adopted here is consistent with that commonly used in observations. 

One can show that the derivation of $\Phi_{\rm B}$ can be written as:
\begin{equation}
    \frac{d\Phi_{\rm B}}{ds} \; = \;  F_\varphi \frac{\Omega}{c} = F_\varphi \frac{1}{r_l} 
    \quad \hbox{with} \quad
     F_\varphi \; = \;  \frac{-\cos\Phi \cos \alpha \sin \alpha \sin \zeta +\cos \zeta \sin^2 \alpha }{1 - \left(\cos\alpha \cos \zeta + \sin\alpha \sin\zeta \cos\Phi\right)^2}  
     \label{eq:dphiB_ds_Fphi}
\end{equation}
%
%
%
%
Substituting Equation~(\ref{eq:dphiB_ds_Fphi}) into Equation~
(\ref{eq:app-adiabatic_rpl_AL}), one obtains
\begin{align}
     k\Delta n(r = r_{\rm rec}) \; = \;  2F_\varphi \frac{1}{r_l} \quad . \label{eq:rpl_AL}
\end{align}
%
Using Equations~(\ref{eq:deltan}) and
 (\ref{eq:theta_AL}), Equation~(\ref{eq:rpl_AL}) becomes:
\begin{equation}
    \left(\frac{r_{\rm rec}}{R_{\rm NS}}\right)^6 \; = \; r_l\frac{\alpha_{\rm f}}{120}\frac{\nu}{c}\left(\frac{B_p}{B_{\rm cr}}\right)^2 \frac{F_B}{F_\varphi} \quad ,
\end{equation}
where $ F_B = 1 -  \left(\cos\alpha \cos \zeta + \sin\alpha \sin\zeta \cos\Phi\right)^2 = 1 - \cos^2 \theta = \sin^2\theta$.
Using  $E_1 = {E_{\gamma}} / ({1\mathrm{keV}})$, \edithere{$P_1 = P / (1s)$, with $P$ being the spin period of the NS}, and $B_{12} = {B_p}/{(10^{12} G)}$, the recoupling radius can be written as
\begin{equation}
    \frac{r_{\rm rec}}{R_{\rm NS}} \; \approx \; 32.6 \left(\frac{\edithere{P_1} E_1 B^2_{12}F_B}{F_{\varphi}}\right)^{1/6} \quad ,
\end{equation}
which is the same as Equation~(67) in \citet{Adelsberg-2006-MNRAS}, and is equivalent to Equation~(\ref{eq:rplAP_final}).


\newpage

\bibliography{ref}{}
\bibliographystyle{aasjournal}



\end{document}